\documentclass[11pt,letterpaper]{JHEP3}
\usepackage{amsmath,epsfig,array,undertilde}
\usepackage{graphicx}
\usepackage{amssymb}
\usepackage{amsfonts,amstext,latexsym,xspace}
\usepackage{amsfonts}

\newcommand{\eq}{\begin{equation}}
\newcommand{\en}{\end{equation}}
\newcommand{\eqn}{\begin{eqnarray}}
\newcommand{\enn}{\end{eqnarray}}
\newcommand{\nn}{\nonumber }
\newcommand{\beq}{\begin{equation}}
\newcommand{\eeq}{\end{equation}}
\newcommand{\M}{\ensuremath{\mathcal{M}}}

\newcommand{\gd}[1][{}]{\delta_{#1}{}}

\newcommand{\cX}{\mathcal{X}}
\newcommand{\cY}{\mathcal{Y}}

\newcommand{\cN}{\mathcal{N}}

\newcommand{\FTP}[4][{}]{\left( #2,#3,#4 \right)^{#1}}

\title{Spectrum Generating Conformal and  Quasiconformal U-Duality Groups, Supergravity and Spherical Vectors}
\author{
Murat G\"{u}naydin$^{\dagger}$\footnote{murat@phys.psu.edu}
\\
$^{\dagger}$ \emph{Institute for Gravitation and the Cosmos \\ Physics Department \\
Penn  State University\\
University Park, PA 16802, USA} \\
}
\author{Oleksandr Pavlyk$^{\ddagger}$\footnote{pavlyk@wolfram.com}
\\
$^{\ddagger}$\emph{Wolfram Research Inc. \\100 Trade Center Dr. \\Champaign, IL 61820, USA} }

\abstract{
After reviewing the algebraic structures that underlie the geometries
 of $N=2$ Maxwell-Einstein supergravity theories (MESGT) in five and
 four dimensions with symmetric scalar manifolds, we give a unified
 realization of their three dimensional U-duality groups as spectrum
 generating quasiconformal groups. They are $F_{4(4)}, E_{6(2)},
 E_{7(-5)}, E_{8(-24)}$ and $SO(n+2,4)$.  Our formulation is covariant
 with respect to U-duality symmetry groups of corresponding five
 dimensional supergravity theories, which are $SL(3,\mathbb{R})$,
 $SL(3,\mathbb{C})$, $SU^*(6)$, $E_{6(6)}$ and $SO(n-1,1)\times SO(1,1)$,
 respectively. We determine the spherical vectors of quasiconformal
 realizations of all these groups twisted by a unitary character. We also give their quadratic Casimir operators and determine their values. Our
 work lays the algebraic groundwork for constructing the unitary
 representations of these groups induced by their geometric
 quasiconformal actions, which include the quaternionic discrete
 series. For rank 2 cases, $SU(2,1)$ and $G_{2(2)}$, corresponding to
 simple $N=2$ supergravity in four and five dimensions, this program
 was carried out in arXiv:0707.1669.  We also discuss the corresponding  algebraic
 structures underlying symmetries of $N\geq 4$ supergravity theories.
 They lead to quasiconformal realizations of split real forms of
 U-duality groups as a straightforward extension of the quaternionic
 real forms. }

\preprint{}
\begin{document}
\renewcommand{\theequation}{\arabic{section}.{\arabic{subsection}}.\arabic{equation}}
\section{Introduction}
\setcounter{equation}{0}
Earliest studies of unitary representations of U-duality groups of
extended supergravity theories were given in
\cite{Gunaydin:1981dc,Gunaydin:1981yq,Gunaydin:1981zm}. These studies
were motivated mainly by the idea that in a quantum theory global
symmetries must be realized unitarily on the spectrum as well as by
the composite scenarios that attempted to connect maximal $N=8$
supergravity with observation \cite{Ellis:1980tf,Ellis:1980cf,Gunaydin:1982gw}\footnote{ For further references on the subject, see \cite{Gunaydin:1982gw}} . In
the composite scenario of \cite{Ellis:1980tf} $SU(8)$ local symmetry
of $N=8$ supergravity was conjectured to become dynamical and act as a
family unifying grand unified theory (GUT) which contains $SU(5)$ GUT
as well as a family group $SU(3)$.  A similar scenario leads to $E_6$
GUT with a family group $U(1)$ in the exceptional supergravity theory
\cite{Gunaydin:1983rk}. With the discovery of counter terms at higher
loops suggesting that divergences at higher orders would eventually spoil the
finiteness properties of $N=8$ supergravity and the discovery of
Green-Schwarz anomaly cancellation in superstring theory \cite{Green:1984sg}, the attempts
at composite scenarios were abandoned. However, recent discovery of
unexpected cancellations of divergences in supergravity theories
\cite{Bern:2008pv,BjerrumBohr:2008dp,ArkaniHamed:2008gz,Chalmers:2000ks,Green:2006gt,Green:2006yu,Green:2007zzb,Kallosh:2008ru} has brought back the question of finiteness of
$N=8$ supergravity as well as of exceptional supergravity.

Over the last decade or so, work on unitary representations of
U-duality groups of extended supergravity theories has accelerated,
leading to considerable progress. Some of the renewed interest in
unitary realizations of U-duality groups was due to the proposals that
certain extensions of U-duality groups may act as spectrum generating
symmetry groups. First, based on geometric considerations involving
orbits of extremal black hole solutions in $N=8$ supergravity and
$N=2$ Maxwell-Einstein supergravity theories with symmetric scalar
manifolds, it was suggested that four dimensional U-duality groups act
as spectrum generating conformal symmetry groups of the corresponding
five dimensional supergravity theories
\cite{Ferrara:1997uz,Gunaydin:2000xr,Gunaydin:2004ku,Gunaydin:2003qm,Gunaydin:2005gd}. This
proposal was extended to the proposal that three dimensional U-duality
groups act as spectrum generating quasiconformal groups of the
corresponding four dimensional supergravity theories
\cite{Gunaydin:2000xr,Gunaydin:2004ku,Gunaydin:2003qm,Gunaydin:2005gd}. Quasiconformal
realization of the U-duality group $E_{8(8)}$ of maximal supergravity
in three dimension is the first known geometric realization of
$E_{8(8)}$ and its quantization leads to the minimal unitary representation of $E_{8(8)}$ \cite{Gunaydin:2001bt}.  Remarkably, quasiconformal realizations exist for
different real forms of all noncompact groups and their quantizations
yield directly  the minimal unitary representations of the
respective groups
\cite{Gunaydin:2001bt,Gunaydin:2004md,Gunaydin:2005zz,Gunaydin:2006vz}. In
fact, the construction of minimal unitary representations of
noncompact groups via the quasiconformal method gives a unified
approach to these representations and extends also to supergroups
\cite{Gunaydin:2006vz}. For symplectic groups these minimal unitary
representations are simply the singleton representations.

Five dimensional Maxwell-Einstein supergravity theories with symmetric
scalar manifolds $\mathcal{M}_5=G_5/K_5$ such that $G_5$ is a symmetry
of the Lagrangian are in one-to-one correspondence with Euclidean
Jordan algebras $J$ of degree three and their scalar manifolds are of
the form
\begin{equation*}
  \mathcal{M}_5 = \frac{Str_0(J)}{Aut(J)}
\end{equation*}
where $Str_0(J)$ and $Aut(J)$ are the reduced structure and
automorphism groups of the Jordan algebra $J$, respectively.  The
scalar manifolds of these theories in four dimensions are
\begin{equation*}
 \mathcal{M}_4 = \frac{Conf(J)}{ \widetilde{Str}_0 (J) \times U(1)}
\end{equation*}
where $Conf(J)$ is the conformal group of the Jordan algebra $J$ and
$\widetilde{Str}_0(J)$ is the compact form of the reduced structure
group. Upon further dimensional reduction to three dimensions they
lead to scalar manifolds of the form
\begin{equation*}
 \mathcal{M}_3 = \frac{QConf(J)}{\widetilde{Conf}(J) \times SU(2)}
\end{equation*}
where $QConf(J)$ is the quasiconformal group associated with the
Jordan algebra $J$ and $\widetilde{Conf}(J)$ is the compact form of
the conformal group of $J$.

Since they were first proposed many results have been obtained that
support the proposals that four and three dimensional U-duality groups
act as spectrum generating conformal and quasiconformal groups of five
and four dimensional supergravity theories with symmetric scalar
manifolds. First, the work relating black hole solutions in four and
five dimensions
\cite{Gaiotto:2005xt,Gaiotto:2005gf,Elvang:2005sa,Pioline:2005vi} is
in complete accord with the proposal that four dimensional U-duality
groups act as spectrum generating conformal symmetry groups of five
dimensional supergravity theories from which they
descend. Furthermore, applying the two proposals in succession leads
to the proposal that three dimensional U-duality groups should act as
spectrum generating symmetry groups of five dimensional supergravity
theories. The work of \cite{Bouchareb:2007ax,Gal'tsov:2008nz} on using
solution generating techniques to relate the known black hole
solutions of five dimensional ungauged supergravity theories to each other and
generate new solutions using symmetry groups of the corresponding
three dimensional supergravity theories and related work on gauged supergravity theories \cite{Berkooz:2008rj} lend further support to these
proposals.

A concrete framework for implementation of the proposal that three
dimensional U-duality groups act as spectrum generating quasiconformal
groups was given in
\cite{Gunaydin:2005mx,Gunaydin:2007bg,Gunaydin:2007qq} for spherically
symmetric stationary BPS black holes of four dimensional supergravity
theories. The basic starting point is the observation that the
attractor equations \cite{Ferrara:1995ih,Ferrara:1996um} symmetric stationary black holes of
four dimensional supergravity theories are equivalent to the geodesic
motion of a fiducial particle on the moduli space $\mathcal{M}^*_3$ of
the three dimensional supergravity obtained by reduction on a
time-like circle\footnote{ This was first observed in
\cite{Breitenlohner:1987dg} and used in
\cite{Cvetic:1995kv,Cvetic:1995uj} to construct static and rotating
black holes in heterotic string theory.}. A related analysis on
non-BPS extremal black holes in theories with symmetric target
manifolds was carried out in \cite{Gaiotto:2007ag}. For $N=2$ MESGTs
defined by Jordan algebras of degree three the manifolds
$\mathcal{M}^*_3$ are para-quaternionic symmetric spaces
\begin{equation*}
 \mathcal{M}^*_3 =\frac{QConf(J)} { Conf(J)\times SU(1,1) }
\end{equation*}
Quantization of the motion of fiducial particle leads to quantum
mechanical wave functions that provide the basis of a unitary
representation of $QConf(J)$.  BPS black holes correspond to a special
class of geodesics which lift holomorphically to the twistor space
$\mathcal{Z}_3$ of $\mathcal{M}_3^*$ and $\mathcal{Z}_3$ can be
identified as the BPS phase space. Spherically symmetric stationary
BPS black holes of $N=2$ MESGT's are described by holomorphic curves
in $\mathcal{Z}_3$
\cite{Gunaydin:2005mx,Gunaydin:2007bg,Gunaydin:2007qq,Neitzke:2007ke}.
The quasiconformal group actions discovered in \cite{Gunaydin:2000xr}
extend to the complexified groups. As a consequence, one finds that
the action of three dimensional U-duality group $QConf(J)$ on the
natural complex coordinates of corresponding twistor space is
precisely of the quasiconformal type \cite{Gunaydin:2007qq}.
Therefore the unitary representations of $QConf(J)$ relevant for BPS
black holes are those induced by holomorphic quasiconformal actions of
$QConf(J)$ on the corresponding twistor spaces $\mathcal{Z}_3$, which
belong in general to quaternionic discrete series representations \cite{Gunaydin:2007qq}.

Further support for the proposal that three dimensional U-duality
groups act as spectrum generating groups of the corresponding four
dimensional theories comes from the connection established in
\cite{Gunaydin:2007vc} between the harmonic superspace (HSS)
formulation of $N=2$, $d=4$ supersymmetric quaternionic K\"ahler sigma
models that couple to $N=2$ supergravity and the minimal unitary
representations of their isometry groups. In particular, for $N=2$
sigma models with quaternionic symmetric target spaces of the form
$QConf(J)/\widetilde{Conf}(J)\times SU(2)$ one finds that there exists
a one-to-one mapping between the quartic Killing potentials that
generate the isometry group $ QConf(J)$ under Poisson brackets in the
HSS formulation, and the generators of minimal unitary representation
of $ QConf(J)$.  This suggests that in the corresponding quantum
theory the ``fundamental spectrum'' fits into the minimal unitary
representation of $QConf(J)$ and the full spectrum is obtained by
tensoring of the minimal unitary representation obtained from
quantization of the quasiconformal action. We should perhaps stress
that the results of \cite{ Gunaydin:2007vc} apply to all quaternionic
K\"ahler sigma models and not only those that are in the C-map of four
dimensional $N=2$ MESGT's.

In \cite{Gunaydin:2007qq} unitary representations of two quaternionic
groups of rank two, namely $SU(2,1)$ and $G_{2(2)}$, induced by
their geometric quasiconformal actions were studied in great
detail. These groups are the isometry groups of four and five
dimensional simple $N=2$ supergravity theories  dimensionally reduced to three
dimensions, respectively.  Unitary representations induced by the
quasiconformal action twisted by a unitary character include the
quaternionic discrete series representations that were studied in
mathematics literature using algebraic methods \cite{MR1421947}.  The
starting point of the constructions of unitary representations of
$SU(2,1)$ and $G_{2(2)}$ given in \cite{Gunaydin:2007qq} are the spherical
vectors of maximal compact subgroups under their quasiconformal
actions. In this paper we lay the algebraic groundwork for extending
the results of \cite{Gunaydin:2007qq} to all quaternionic
quasiconformal groups $QConf(J)$ that arise as three dimensional
U-duality groups of five dimensional $N=2$ MESGT's defined by
Euclidean Jordan algebras $J$ of degree three. We also present some
results for extending this program to quasiconformal groups associated
with non-Euclidean Jordan algebras of degree three that correspond to
three dimensional U-duality groups of supergravity theories with $N \geq 4$ supersymmetries
in $d=4$ or $d=5$. Study of the unitary representations induced from the quasiconformal actions on the spherical vectors will be the subjects of separate studies.

In section 2 we review the five and four dimensional $N=2$ MESGT's
with symmetric scalar manifolds and the algebraic structures that
underlie their geometries, namely, Euclidean Jordan of degree three
and Freudenthal triple systems defined over them,
respectively. Section 3 reviews the symmetries of supergravity
theories with $N \geq 4$ in five, four and three dimensions and
discuss their relation to non-Euclidean Jordan algebras of degree
three and associated Freudenthal triple systems. In section 4 we give
a unified realization of all quaternionic quasiconformal groups
$QConf(J)$ in a basis covariant with respect to their five dimensional
U-duality groups, which are simply the reduced structures groups of
Euclidean Jordan algebras $J$ and and identify the generators of their
maximal compact subgroups. Again in section 4 we present the quadratic Casimir operators of all the quasiconformal 
algebras $QConf(J)$ and determine their values.  Section 5 is devoted to the study of the
quasiconformal group $SO(4,4)$ of the Jordan algebra $ J = \mathbb{R}
\oplus \mathbb{R} \oplus \mathbb{R} $ coresponding to the direct sum
of three irreducible idempotents. Every element of a Euclidean Jordan
algebra of degree three can be brought to this form by the action of
its automorphism group and hence study of $SO(4,4)$ as a
quasiconformal group is important. The corresponding supergravity
theory is the STU model. We give the commutation relations in a
noncompact basis covariant with respect to the reduced structure group
$ SO(1,1)\times SO(1,1)$ as well as in a compact basis. We give the
spherical vector of quasiconformal $SO(4,4)$ annihilated by all the
generators of its maximal compact subgroup $SO(4)\times SO(4) =
SU(2)\times SU(2) \times SU(2) \times SU(2) $. In section 6 we
determine the spherical vectors of all quaternionic quasiconformal
groups $QConf(J)$ annihilated by the generators of their maximal
compact subgroups $\widetilde{Conf}(J) \times SU(2)$. The
quasiconformal groups associated with simple Euclidean Jordan algebras
are $F_{4(4)}, E_{6(2)}, E_{7(-5)}$ and $E_{8(-24)}$ and those
associated with nonsimple Euclidean Jordan algebras are
$SO(n+2,4)$. In section 7 we briefly discuss how the results of
section 4 extend to split exceptional groups $E_{6(6)}, E_{7(7)}$ and
$E_{8(8)}$ and to $SO(m+4,n+4)$ by simply replacing the Euclidean
Jordan algebras by split Jordan algebras. The appendix A gives a
review of relevant Jordan algebra theory. Appendix B reviews the
construction of conformal algebras of Jordan algebras and Appendix C
reviews the quasiconformal group actions over Freudenthal triple
systems extended by an extra coordinate.

\section{Conformal and Quasiconformal Groups of Jordan Algebras and Maxwell-Einstein Supergravity Theories}
\setcounter{equation}{0}
Five dimensional Maxwell-Einstein supergravity theories with symmetric
scalar manifolds $G/H$ such that $G$ is a symmetry of the Lagrangian
are uniquely defined and classified by Euclidean Jordan algebras of
degree three. The corresponding four dimensional MESGTs obtained by
dimensional reduction are similarly described by Freudenthal triple
systems defined over these Jordan algebras.  These triple systems were
introduced by Freudenthal in his study of metaplectic geometries
associated with the exceptional groups $F_4, E_6, E_7$ and $E_8$
\cite{MR0170974,MR0063358}. They correspond to the last row of some very
remarkable  geometries associated with the groups of the Magic
Square. Referring to appendices A,B and C for further details on
Jordan algebras, Freudenthal triple systems and their symmetry groups
we shall in this section review these MESGTs and associated algebraic
and geometric structures.

\subsection{5D, $N=2$ Maxwell-Einstein Supergravity Theories  and Jordan Algebras}
\setcounter{equation}{0}
Five dimensional MESGTs that describe the coupling of an arbitrary number
of $N=2$ (Abelian) vector multiplets to $N=2$ supergravity were constructed
in \cite{Gunaydin:1983rk,Gunaydin:1983bi,Gunaydin:1984ak,Gunaydin:1986fg}.
The bosonic part of five dimensional  $N=2$ MESGT Lagrangian is given by\footnote{%
We use the conventions of \cite{Gunaydin:1983bi} in this section.}
\begin{eqnarray}\label{Lagrange}
e^{-1}\mathcal{L}_\textrm{bosonic} &=& -\frac{1}{2} R
-\frac{1}{4}{\stackrel{\circ}{a}}_{IJ} F_{\mu\nu}^{I} F^{J\mu\nu}-
 \frac{1}{2} g_{xy}(\partial_{\mu}\varphi^{x})
(\partial^{\mu} \varphi^{y})+\nonumber \\
 && + \frac{e^{-1}}{6\sqrt{6}} C_{IJK} \varepsilon^{\mu\nu\rho\sigma\lambda}
 F_{\mu\nu}^{I}F_{\rho\sigma}^{J}A_{\lambda}^{K},
\end{eqnarray}
where
\begin{eqnarray*}
I&=& 1,\ldots, n_V\\
a&=& 1,\ldots, (n_V-1)\\
x&=& 1,\ldots, (n_V-1)\\
\mu, \nu,... &=& 0,1,2,3,4.
\end{eqnarray*}
$e$ and $R$, respectively, denote the f\"{u}nfbein determinant and
scalar curvature of spacetime. $F_{\mu\nu}^{I}$ are field strengths
of the vector fields $A_{\mu}^{I}$. The metric, $g_{xy}$, of the
scalar manifold $\M_5$ and the metric ${\stackrel{\circ}{a}}_{IJ}$
both depend on the scalar fields $\varphi^{x}$. On the other hand, the
completely symmetric tensor $C_{IJK}$ is constant as required by local
Abelian gauge symmetries of vector fields.

One remarkable feature of these theories is the fact that the entire
$N=2$ MESGT is uniquely determined by the constant tensor $C_{IJK}$
\cite{Gunaydin:1983bi}. In particular, geometry of the scalar manifold
$\mathcal{M}_5$ is determined by $C_{IJK}$, which can be used to
define a cubic polynomial, $\mathcal{V}(h)$, in $n_V$ real variables
$h^{I}$ $(I=1,\ldots,n_V)$,
\begin{equation}
 \mathcal{V}(h):=C_{IJK} h^{I} h^{J} h^{K}\ .
\end{equation}
One defines a metric, $a_{IJ}$, of an ambient space
$\mathcal{C}_{n_V}$ coordinatized by $h^{I}$:
\begin{equation}\label{aij}
  a_{IJ}(h):=-\frac{1}{3}\frac{\partial}{\partial h^{I}}
  \frac{\partial}{\partial h^{J}} \ln \mathcal{V}(h)\ .
\end{equation}
and $(n_V-1)$-dimensional manifold, $\mathcal{M}_5$, of scalar fields
$\varphi^{x}$ can then be represented as an hypersurface defined by
the condition \cite{Gunaydin:1983bi}
\begin{equation}\label{hyper1}
{\cal V} (h)=C_{IJK}h^{I}h^{J}h^{K}=1 \ ,
\end{equation}
in this ambient space $\mathcal{C}_{n_V}$. The ambient space
$\mathcal{C}_{n_V}$ is the 5 dimensional counterpart of the hyper
K\"ahler cone of the twistor space of the corresponding three
dimensional quaternionic geometry of the scalar manifold
$\mathcal{M}_3$.  The metric $g_{xy}$ is simply the pull-back of
(\ref{aij}) to $\mathcal{M}_5$:
\begin{equation}
g_{xy}(\varphi)= h^{I}_{x}  h^{J}_{y} a_{IJ}|_{\mathcal{V}=1}
\end{equation}
where $ h^{I}_{x} = \sqrt{\frac{3}{2}}
\frac{\partial}{\partial\phi^{x}} h^{I} $ and the ``metric''
${\stackrel{\circ}{a}}_{IJ}(\varphi)$ of kinetic energy term of the
vector fields is given by the componentwise restriction of the metric
$a_{IJ}$ of the ambient space $\mathcal{C}_{n_V}$ to $\mathcal{M}_5$:
\begin{equation*}
{\stackrel{\circ}{a}}_{IJ}(\varphi)=a_{IJ}|_{{\cal V}=1} \ .
\end{equation*}
Riemann curvature tensor of the scalar manifold has a very simple form
\begin{equation}
  K_{xyzu}= \frac{4}{3} \left( g_{x[u} g_{z]y} + {T_{x[u}}^{w} T_{z]yw} \right)
\end{equation}
where $T_{xyz}$ is the symmetric tensor
\begin{equation}
   T_{xyz}= h^{I}_{x }h^{J}_{y} h^{K}_{z} C_{IJK}
\end{equation}
From the form of the Riemann curvature tensor $K_{xyzu}$ it follows
that the covariant constancy of $T_{xyz}$ implies the covariant
constancy of $K_{xyzu}$:
\begin{equation*}
T_{xyz ; w} = 0
\longrightarrow
 K_{xyzu ; w} =0
\end{equation*}
Therefore scalar manifolds $\mathcal{M}_5$ with covariantly constant
$T$ tensors are locally symmetric spaces.

If $\mathcal{M}_5$ is a homogeneous space the covariant constancy of
$T_{xyz}$ was shown to be equivalent to the ``adjoint
identity''~\cite{Gunaydin:1983bi}:
\begin{equation}
   C^{IJK} C_{J(MN} C_{PQ)K} = \delta^{I}_{(M} C_{NPQ)}
\end{equation}
where the indices are raised by the inverse
${\stackrel{\circ}{a}}{}^{IJ}$ of ${\stackrel{\circ}{a}}{}_{IJ}$.
Furthermore, cubic forms defined by $C_{IJK}$ of $N=2$ MESGT's that
satisfy the adjoint identity are in one-to-one correspondence with
norm forms of Euclidean (formally real) Jordan algebras of degree 3
\cite{Gunaydin:1983bi}.  These theories exhaust the list of $5D$
MESGTs with symmetric target spaces $G/H$ such that $G$ is a symmetry
of their Lagrangians \cite{deWit:1992wf}.  Remarkably, the list of
cubic forms that satisfy the adjoint identity coincides with the list
of Legendre invariant cubic forms that were classified more recently
\cite{MR2094111}.

Symmetric target spaces of $N=2$ MESGTs defined by Euclidean Jordan
algebras of degree three are of the form
\begin{equation}
    \mathcal{M} = \frac{\mathrm{Str}_0 \left(J\right)}{ \mathrm{Aut}\left(J\right)}
\end{equation}
where $\mathrm{Str}_0\left(J\right)$ and $\mathrm{Aut}\left(J\right)$
are the reduced structure (``Lorentz'') group and automorphism
(``rotation'') group of the Jordan algebra $J$ respectively.  Vector
fields of these theories are in one-to-one correspondence with
elements of the underlying Jordan algebra $J$ and transform linearly
under the action of $\mathrm{Str}_0 (J)$.  Similarly, charges to which
vector fields couple transform linearly under the action of $
\mathrm{Str}_0 (J)$, which can be interpreted as the ``Lorentz'' group
of the Jordan algebra.  Therefore, to a black hole solution of the
corresponding MESGTs with charges $q^{I}$ one can associate an element
of the underlying Jordan algebra $q=e_{I}q^{I}$, where $e_{I}$ form a
basis of $J$. The entropy of a spherically symmetric stationary
extremal black hole solution is then determined by the cubic norm
$\mathcal{V}(q)$ of $q$. The orbits of spherically symmetric
stationary extremal black holes of five dimensional $N=2$ MESGT's
defined by Jordan algebras were first classified in
\cite{Ferrara:1997uz}. It was found that the solutions with vanishing
entropy have larger symmetries beyond those of the reduced structure
group, namely they are invariant under ``special conformal
transformations'' of the Jordan algebra $J$ that lie outside the
Lorentz group of $J$.  As reviewed in Appendix B one can define the
action of a generalized conformal group $Conf(J)$ acting on the Jordan
algebra $J$ \cite{Gunaydin:1975mp,Gunaydin:1992zh,Gunaydin:2000xr},
which leaves light-like separations $\mathcal{V}\left(q-q'\right)=0$
invariant. Conformal group $Conf(J)$ changes the norm of a general
element $q\in J$ and hence the corresponding entropy of black hole
solutions. Therefore conformal groups $Conf(J)$ of Jordan algebras
that define $N=2$ MESGT's were proposed as spectrum generating
symmetry groups of the solutions of these theories in five dimensions
\cite{Ferrara:1997uz,Gunaydin:2000xr,Gunaydin:2003qm,Gunaydin:2004ku,Gunaydin:2005gd}.
Furthermore, $Conf(J)$ is isomorphic to the U-duality symmetry group
of the corresponding four dimensional MESGT obtained by dimensional
reduction (R-map).

\subsection{4D, $N=2$  Maxwell-Einstein Supergravity Theories and Freudenthal Triple Systems}
\setcounter{equation}{0}
Let us now briefly review the dimensional reduction of the 5D, $N=2$
MESGTs to four dimensions (R-map), restricting ourselves to the
bosonic sector. The metric of the target space of the four-dimensional
scalar fields of dimensionally reduced theories were first obtained in
\cite{Gunaydin:1983bi}, where it was shown that these four
dimensional target spaces are generalized upper half-spaces (tube
domains) over the convex cones defined by the cubic norm. More
specifically, the four dimensional scalar manifold are parametrized by
complex coordinates \cite{Gunaydin:1983bi},
\begin{equation}
   z^{I}:=\frac{1}{\sqrt{3}}A^{I} + \frac{i}{\sqrt{2}}\tilde{h}^{I}
\end{equation}
where $A^{I}$ denote the 4d scalars descending from the 5D vectors. Imaginary components are given by
\begin{equation}
   \tilde{h}^{I}:=e^\sigma h^{I}. \label{htildeI}
\end{equation}
where $\sigma$ is the scalar field coming from 5D graviton and they
satisfy the positivity condition
\begin{equation*}
  C_{IJK} \tilde{h}^{I}\tilde{h}^{J}\tilde{h}^{K}= e^{3\sigma} >0
\end{equation*}
Geometry of four dimensional $N=2$ MESGTs obtained by dimensional
reduction from five dimensions (R-map) was later called ``very special
geometry'' and has been studied extensively \cite{deWit:1991nm}.  Dimensional reduction of
the full bosonic sector of gauged $5D$ $N=2$ MESGTs with tensor
multiplets and its reformulation in the language of special K\"ahler
geometry was given in \cite{Gunaydin:2005bf}, which we follow in our
summary here, restricting ourselves to the ungauged theory.

The $n_V$ complex coordinates $z^{I}$ can be interpreted as
inhomogeneous coordinates of a $(n_V+1)$-dimensional complex vector
\begin{equation}
 X^A=  \left( \begin{array}{c} X^{0}\\ X^{I} \end{array} \right) =
       \left( \begin{array}{c}    1 \\ z^{I} \end{array} \right) .
\end{equation}
Taking as ``prepotential'' the cubic form
\begin{equation}
  F(X^A)=-\frac{\sqrt{2}}{3} C_{IJK}\frac{X^{I}X^{J}X^{K}}{X^{0}}
\label{prepot}
\end{equation}
and using the symplectic section
\begin{equation}
  \left( \begin{array}{c} X^A\\ F_{A}   \end{array} \right)  =
  \left( \begin{array}{c} X^A\\ \partial_{A}F \end{array} \right) \equiv
 \left( \begin{array}{c}  X^A\\ \frac{\partial F}{\partial X^{A}} \end{array}
\right)
\end{equation}
one obtains the K\"{a}hler potential
\begin{eqnarray}
K(X,\bar{X})&:=&-\ln [i\bar{X}^{A}F_{A}-iX^{A}\bar{F}_{A}] \label{symK}\\
&=&-\ln  \left[  i\frac{\sqrt{2}}{3}C_{IJK}(z^{I}-\bar{z}^{I})(z^{J}-\bar{z}^{J})(z^{K}-\bar{z}^{K})    \right] \label{kahlerpotential}
\end{eqnarray}
and the ``period matrix''
\begin{equation}
\mathcal{N}_{AB}:=\bar{F}_{AB}+2i\frac{\textrm{Im}(F_{AC})\textrm{Im}(F_{BD})X^{C}X^{D}}{\textrm{Im}(F_{CD})X^{C}X^{D}}
\end{equation}
where $F_{AB}\equiv \partial_{A}\partial_{B}F$ etc.  Components of the
resulting period matrix $\mathcal{N}_{AB}$ are:
\begin{eqnarray}
\mathcal{N}_{00}&=&-\frac{2\sqrt{2}}{9\sqrt{3}}C_{IJK}A^{I}A^{J}A^{K}
-\frac{i}{3} \left( e^{\sigma}{\stackrel{\circ}{a}}_{IJ} A^{I}A^{J}+\frac{1}{2} e^{3\sigma} \right)\\
\mathcal{N}_{0I}&=&\frac{\sqrt{2}}{3}C_{IJK}A^{J}A^{K}+\frac{i}{\sqrt{3}} e^{\sigma} {\stackrel{\circ}{a}}_{IJ} A^{J}\\
\mathcal{N}_{IJ}&=&  -\frac{2\sqrt{2}}{\sqrt{3}}C_{IJK}A^{K}-ie^{\sigma}
{\stackrel{\circ}{a}}_{IJ} \label{NIJ}
\end{eqnarray}
Prepotential (\ref{prepot}) leads to the K\"ahler metric
\begin{equation}\label{metric}
  g_{I\bar{J}}  \equiv \partial_{I}\partial_{\bar{J}}K = \frac{3}{2} e^{-2\sigma} {\stackrel{\circ}{a}}_{IJ}
\end{equation}
for the scalar manifold $\M_{(4)}$ of four-dimensional theory.
Denoting the field strength of vector field that comes from the
graviton in five dimensions as
$F_{\mu\nu}^{0}$
 bosonic sector of dimensionally reduced Lagrangian can be written as
\begin{eqnarray}
   e^{-1}\mathcal{L}^{(4)} &=&-\frac{1}{2}R  -
   g_{I\bar{J}}    (\partial_{\mu}z^{I})(\partial^{\mu} \bar{z}^{J})
   \nonumber \\
   &&+\frac{1}{4}\textrm{Im}(\mathcal{N}_{AB})F_{\mu\nu}^{A}F^{\mu\nu B}-\frac{1}{8}
   \textrm{Re} (\mathcal{N}_{AB})\epsilon^{\mu\nu\rho\sigma}
   F_{\mu\nu}^{A}F_{\rho\sigma}^{B}.\label{redlag1b}
\end{eqnarray}

Since scalar fields $z^{I}$ are restricted to the domain $
\mathcal{V}(Im(z)) > 0$, the scalar manifolds of 4D, $N=2$ MESGT's
defined by Euclidean Jordan algebras $J$ of degree three are simply
the K\"ocher ``upper half spaces'' of these Jordan algebras, which
belong to the family of Siegel domains of the first kind
\cite{MR1718170}.  The ``upper half spaces'' of Jordan algebras can be
mapped into bounded symmetric domains, which can be realized as
hermitian symmetric spaces of the form
\begin{equation}
  \mathcal{M}_4= \frac{Conf(J)}{\widetilde{Str}{J}}
\end{equation}
where maximal compact subgroup $\widetilde{Str}{J}$ of the conformal
group of $J$ is the compact real form of the structure group $Str(J)$
generated by dilatations and Lorentz transformations of $J$.  We
should also stress the important point that the K\"ahler potential
\ref{kahlerpotential} that one obtains directly under dimensional
reduction from 5 dimensions is given by the ``cubic light-cone''
\begin{equation}
  \mathcal{V}(z-\bar{z}) =C_{IJK}(z^{I}-\bar{z}^{I})(z^{J}-\bar{z}^{J})(z^{K}-\bar{z}^{K})
\end{equation}
which is manifestly invariant under the 5 dimensional U-duality group
$Str_0(J)$ and real translations
\begin{equation*}
 Re (z^I)\Rightarrow Re(z^I) + a^I
\end{equation*}
\begin{equation*}
  a^I \in \mathbb{R}
\end{equation*}
which follows from Abelian gauge invariances of vector fields of the
five dimensional theory. Under dilatations it gets simply
rescaled. Infinitesimal action of special conformal generators $K^{I}$
on the ``cubic light-cone'' gives \cite{Gunaydin:2000xr}
\begin{equation}
   K^{I} \mathcal{V}(z-\bar{z}) = (z^{I} + \bar{z}^{I}) \mathcal{V}(z-\bar{z})
\end{equation}
which can be integrated to give the global transformation:
\begin{equation}
\mathcal{V}(z-\bar{z}) \Longrightarrow f(z^{I}) \bar{f}( \bar{z}^{I}) \mathcal{V}(z-\bar{z})
\end{equation}
Thus the cubic light-cone defined by $\mathcal{V}(z-\bar{z}) =0$ is
invariant under the full conformal group $Conf(J)$. Furthermore, the
above global transformation leaves the metric $g_{I\bar{J}}$ invariant
since it corresponds simply to a K\"ahler transformation of the
K\"ahler potential $\ln \mathcal{V}(z-\bar{z}) $.

In $N=2$ MESGTs defined by Euclidean Jordan algebras $J$ of degree
three, one-to-one correspondence between vector fields of five
dimensional theories (and hence their charges) and elements of $J$
gets extended, in four dimensions, to a one-to-one correspondence
between field strengths of vector fields {\it plus} their magnetic
duals and Freudenthal triple systems defined over $J$
\cite{Gunaydin:1983bi,Gunaydin:2000xr,Ferrara:1997uz,Gunaydin:2005gd,Gunaydin:2005zz}. An
element $X$ of Freudenthal triple system (FTS) $\mathcal{F}(J) $ \cite
{MR0170974,MR0063358} over $J$ can be represented formally as a
$2\times 2$ ``matrix'':
\begin{equation}X= \left(
\begin{array}{ccc}
\alpha &  & \mathbf{x} \\
&  &  \\
\mathbf{y} &  & \beta
\end{array}
\right) \in \mathcal{F}(J)
\end{equation}
where $\alpha$, $\beta \in \mathbb{R}$ and $\mathbf{x}$,$\mathbf{y} \in J$

Denoting the ``bare'' four dimensional graviphoton field strength and
its magnetic dual as $F_{\mu \nu }^{0}$ and
$\widetilde{F}_{0}^{\mu\nu}$, respectively, we have the correspondence
\begin{equation*}
\left(
  \begin{array}{ccc}
     F_{\mu \nu }^{0} &  & F_{\mu \nu }^{I} \\ &  &  \\
     \widetilde{F}_{I}^{\mu \nu } &  & \widetilde{F}_{0}^{\mu \nu }
  \end{array}
\right) \Longleftrightarrow
 \left(
    \begin{array}{ccc}
        e_0 &  & e_I \\
            &  &  \\
\tilde{e}^I &  & \tilde{e}^0
    \end{array}
 \right) \in \mathcal{F}(J),
\end{equation*}
where $e_I (\tilde{e}^I)$ are the basis elements of $J$ (its conjugate $\tilde{J}$ ).
Consequently, one can associate with a black hole solution with
electric and magnetic charges (fluxes) $\left(q_0, q_I,
p^{0},p^{I}\right)$ of the 4D MESGT defined by $J$ an element of FTS
$\mathcal{F}\left( J \right) $
\begin{equation}
\left(
\begin{array}{ccc}
p^{0}e_0 &  & p^{I} e_I\\
&  &  \\
q_{I} \tilde{e}^I &  & q_{0}\tilde{e}^0
\end{array}
\right)
\in \mathcal{F}(J) ,
\end{equation}
U-duality group $G_4$ of such a four dimensional MESGT acts as the
automorphism group of FTS $\mathcal{F}(J)$, which is endowed with an
invariant symmetric quartic form and a skew-symmetric bilinear
form. The entropy of a spherically symmetric stationary extremal black with charges $ (p^0, p^I,q_0 ,q_I) $ is determined by the quartic invariant $\mathcal{Q}_4 (q,p)$ of $\mathcal{F}(J)$. With this identification the orbits of extremal black holes of 4D, $N=2$ MESGT's with symmetric scalar manifolds were classified in \cite{Ferrara:1997uz,Bellucci:2006xz}.

As discussed above, four dimensional U-duality groups $G_4$ were
proposed as  spectrum generating conformal symmetry groups in five
dimensions that leave a cubic light-cone invariant. On the other hand,
three dimensional U-duality group $G_3$ of $N=2$ MESGTs defined by
Jordan algebras of degree three do not, in general, have any
conformal realizations on the $2n_V+2$ dimensional space of the FTS.
It was shown in \cite{Gunaydin:2000xr} that three dimensional
U-duality groups $G_3$ have geometric realizations as quasi-conformal
groups on the vector spaces of FTS's  extended by an extra coordinate and
leave invariant a generalized light-cone with respect to a quartic
distance function. This quasiconformal action of three dimensional
U-duality group $G_3$ was proposed as a spectrum generating symmetry
group of corresponding four dimensional supergravity theories
\cite{Gunaydin:2000xr,Gunaydin:2004ku,Gunaydin:2003qm,Gunaydin:2005gd,Gunaydin:2005mx,Gunaydin:2007bg}.
We shall denote the quasicconformal groups defined over FTS's
$\mathcal{F}$ extended by a singlet coordinate as
$QConf(\mathcal{F})$. If the FTS is defined over a Jordan algebra $J$ of degree three we shall denote the corresponding quasiconformal
groups either as $QConf(\mathcal{F}(J))$ or simply as $QConf(J)$.  For
$N=2$ MESGTs defined by Jordan algebras of degree three,
quasiconformal group actions of their three dimensional U-duality
groups $G_3$ were given explicitly in \cite{Gunaydin:2005zz}, in a
basis covariant with respect to U-duality groups of corresponding six
dimensional supergravity theories.

Upon further dimensional reduction to three dimensions (C-map) $N=2$
MESGTs lead to $N=4$, $d=3$ quaternionic K\"ahler $\sigma$ models
coupled to supergravity \cite{Gunaydin:1983bi,Ferrara:1989ik}. In
Table \ref{scalarmanifolds}  we give the symmetry groups of $N=2$ MESGTs defined by Euclidean Jordan
algebras in $d=5,4$ and $3$ dimensions and their scalar manifolds. We
should note that five and three dimensional U-duality symmetry groups
$Str_0(J)$ and $QConf(J)$, respectively, act as symmetries of
supergravity Lagrangians, while four dimensional U-duality groups
$Conf(J)$ are on-shell symmetries.
\begin{table}
\begin{equation} \nonumber
  \begin{array}{|c|c|c|c|} \hline
  ~ &  \mathcal{M}_5 = &  \mathcal{M}_4 = &  \mathcal{M}_3 =  \\
  J & \mathop\mathrm{Str}_0 \left(J\right)/ \mathop\mathrm{Aut}\left(J\right) &
            \mathop\mathrm{Conf} \left(J\right) / \mathop\mathrm{\widetilde{Str}}\left(J\right) &
       \mathop\mathrm{QConf} (\mathcal{F}(J))/
                            \widetilde{ \mathop\mathrm{Conf}\left(J\right)} \times \mathrm{SU}(2) \\[7pt]
        \hline
       J_3^\mathbb{R} & \mathrm{SL}(3, \mathbb{R}) / \mathrm{SO}(3) &
                        \mathrm{Sp}(6, \mathbb{R}) / \mathrm{U}(3)  &
            \mathrm{F}_{4(4)} / \mathrm{USp}(6) \times \mathrm{SU}(2)  \\ [7pt]
       J_3^\mathbb{C} & \mathrm{SL}(3, \mathbb{C}) / \mathrm{SU}(3) &
                        \mathrm{SU}(3, 3) / \mathrm{S}\left(\mathrm{U}(3) \times \mathrm{U}(3)\right) &
            \mathrm{E}_{6(2)} / \mathrm{SU}(6) \times \mathrm{SU}(2)   \\ [7pt]
       J_3^\mathbb{H} & \mathrm{SU}^\ast(6) / \mathrm{USp}(6) &
                        \mathrm{SO}^\ast(12) / \mathrm{U}(6) &
            \mathrm{E}_{7(-5)} / \mathrm{SO}(12) \times \mathrm{SU}(2)  \\[7pt]
       J_3^\mathbb{O} & \mathrm{E}_{6(-26)} / \mathrm{F}_4 &
                        \mathrm{E}_{7(-25)} / \mathrm{E}_6 \times \mathrm{U}(1) &
            \mathrm{E}_{8(-24)} / \mathrm{E}_7 \times \mathrm{SU}(2)  \\[7pt]
       \mathbb{R} \oplus \Gamma_n(\mathbb{Q}) &
                      \frac{  \mathrm{SO}(n-1,1) \times \mathrm{SO}(1,1)}{ \mathrm{SO}(n-1)} &
                       \frac{ \mathrm{SO}(n,2) \times \mathrm{SU}(1,1) }{
                             \mathrm{SO}(n) \times \mathrm{SO}(2) \times \mathrm{U}(1)} &
           \frac{ \mathrm{SO}(n+2,4)}{ \mathrm{SO}(n+2) \times \mathrm{SO}(4)} \\ [7pt]
            \hline
  \end{array}
\end{equation}
\caption{\label{scalarmanifolds}
Above we list the scalar manifolds $\mathcal{M}_d$ of $N=2$ MESGT's
defined by Euclidean Jordan algebras $J$ of degree 3 in $d=3,4,5$
dimensions. $J_3^\mathbb{A}$ denotes the Jordan algebra of $3 \times 3$
Hermitian matrices over the division algebra $\mathbb{A} = \mathbb{R}$,
$\mathbb{C}$, $\mathbb{H}$, $\mathbb{O}$. The last row
$\mathbb{R} \oplus \Gamma_n \left(Q\right)$ are the reducible Jordan
algebras which are direct sums of Jordan algebras $\Gamma_n$ defined
by a quadratic form $\mathbb{Q}$ and a one dimensional Jordan algebra
$\mathbb{R}$. $\mathop\mathrm{\widetilde{Str}}\left(J\right)$ and
$\widetilde{\mathop\mathrm{Conf}}\left(J\right)$ denote the compact
real forms of the structure group $\mathop\mathrm{Str}\left(J\right)$
and conformal group $\mathop\mathrm{Conf}\left(J\right)$ of a Jordan
algebra
$J$. $\mathop\mathrm{QConf}\left(\mathcal{F}\left(J\right)\right)$
denotes the quasiconformal group defined by the FTS
$\mathcal{F}\left(J\right)$ defined over $J$.
}
\end{table}

\section{Symmetries of Supergravity Theories with $N \geq 4$ and non-Euclidean Jordan Algebras of Degree Three}
\renewcommand{\theequation}{\arabic{section}.\arabic{equation}}
\setcounter{equation}{0}
C-tensors and scalar manifolds of simple supergravity theories with
$N>4$ and $N=4$ MESGTs in five spacetime dimensions can also be
described by Jordan algebras of degree three. Referring to appendix A
for further details we summarize the main results in this section. The
C-tensor of five dimensional maximal $N=8$ supergravity can be
identified with the symmetric tensor defining the cubic norm of split
exceptional Jordan algebra $J_3^{\mathbb{O}_s}$ whose invariance group
is $E_{6(6)}$
\cite{Gunaydin:1975mp,Ferrara:1997uz,Gunaydin:2000xr,Gunaydin:2005mx}. Scalar
manifold of $ 5d$, $N=8$ supergravity is
\begin{equation*}
  \mathcal{M}_5 = \frac{E_{6(6)}}{USp(8)}
\end{equation*}
and 27 vector fields of the theory correspond to elements of $
J_3^{\mathbb{O}_s} $, whose reduced structure group is
$E_{6(6)}$. Under dimensional reduction to $d=4$ this correspondence
gets extended to a correspondence between 28 vector field strengths
and their magnetic duals and the Freudenthal triple system
$\mathcal{F}(J_3^{\mathbb{O}_s})$ defined over
$J_3^{\mathbb{O}_s}$. U-duality group $E_{7(7)}$ of four dimensional
$N=8$ supergravity is then simply the automorphism group of
$\mathcal{F}(J_3^{\mathbb{O}_s})$, which is isomorphic to the
conformal group of $J_3^{\mathbb{O}_s}$
\begin{equation}
 Aut(\mathcal{F}(J_3^{\mathbb{O}_s} ) \simeq \mathop\mathrm{Conf}(J_3^{\mathbb{O}_s} )
\end{equation}
U-duality group of 3 dimensional maximal supergravity is $E_{8(8)}$
which is simply the quasiconformal group associated with the
Freudenthal triple system $\mathcal{F}(J_3^{\mathbb{O}_s})$ \cite{Gunaydin:2000xr}
\begin{equation}
  \mathop\mathrm{QConf} (\mathcal{F}(J_3^{\mathbb{O}_s})) = E_{8(8)}
\end{equation}

Remarkably the bosonic sector of $N=6$ supergravity coincide with
bosonic sector of $N=2$ MESGT defined by the Euclidean Jordan algebra
$J_3^{\mathbb{H}}$ and hence their U-duality groups coincide in
$d=5,4$ and 3 dimensions \cite{Gunaydin:1983rk}.  Corresponding scalar
manifolds are given in the third row of Table 1.

As for $N=4$ MESGTs describing the coupling of $n$, $N=4$ vector
multiplets to $N=4$ supergravity, one finds that their C-tensors in
$d=5$ can be identified with non-Euclidean Jordan algebras
\begin{equation}
  (\mathbb{R}\oplus \Gamma_{(5,n)} )
\end{equation}
where $\Gamma_{(5,n)}$ is the Jordan algebra of degree two associated
with a quadratic form of signature $(5,n)$. Scalar manifolds of these
theories in five dimensions are symmetric spaces
\begin{equation*}
\mathcal{M}_5= SO(5,n)\times SO(1,1) /SO(5) \times SO(n)
\end{equation*}
and their vector fields are in one-to-one correspondence with elements
of $(\mathbb{R}\oplus \Gamma_{(5,n)} )$.  In corresponding four
dimensional theories U-duality groups are again automorphism groups of
the underlying Freudenthal triple sytems which are isomorphic to
conformal groups of $(\mathbb{R}\oplus \Gamma_{(5,n)} )$
\begin{equation}
Aut(\mathcal{F}( \mathbb{R}\oplus \Gamma_{(5,n)}  )) \simeq \mathop\mathrm{Conf} (\mathbb{R}\oplus \Gamma_{(5,n)} )
\simeq SO(6,n) \times SU(1,1)
\end{equation}
and scalar manifolds of four dimensional theories are
\begin{equation*}
\mathcal{M}_4= \frac{ SO(6,n+1) \times SU(1,1)}{ SO(6)\times SO(n+1) \times U(1)}
\end{equation*}
In three dimensions their isometry groups are given by  quasiconformal groups of the corresponding Freudenthal triple systems
\begin{equation}
   \mathop\mathrm{QConf} (\mathcal{F}(\mathbb{R}\oplus \Gamma_{(5,n)} )) = SO(8, n+3)
\end{equation}
and their scalar manifolds are simply:
\begin{equation*}
    \mathcal{M}_3  = \frac{ SO(8,n+3)}{SO(8)\times SO(n+3)}
\end{equation*}
In Table \ref{noneuclideantheories} we list the symmetry groups $Aut(J)$, $Str_0(J)$,
$Conf(J)$ and $QConf(J)$ associated with non-Euclidean Jordan algebras
of degree three.
\begin{small}
\begin{table}
\begin{equation} \nonumber
  \begin{array}{|c|c|c|c|c|} \hline
  J & Aut(J) & Str_0(J) & Conf(J) & QConf(J) \\ \hline
  J_3^{\mathbb{C}_s} & SL(3,\mathbb{R}) & SL(3,\mathbb{R}) \times SL(3,\mathbb{R}) & SL(6,\mathbb{R}) & E_{6(6)} \\ \hline
  J_3^{\mathbb{H}_s} & Sp(6,\mathbb{R}) & SL(6,\mathbb{R}) & SO(6,6) &E_{7(7)} \\ \hline
  J_3^{\mathbb{O}_s} & F_{4(4)} & E_{6(6)} & E_{7(7)} & E_{8(8)} \\ \hline
\mathbb{R} \oplus \Gamma_{(n,m)} & SO(n-1,m) & SO(1,1)\times SO(n,m) & SO(n+1,m+1) \times SU(1,1) & SO(n+3,3+m) \\ \hline
\end{array}
\end{equation}
\caption{\label{noneuclideantheories}
Above we give the automorphism ($Aut(J)$), reduced structure
$Str_0(J)$, conformal ($Conf(J)$) and quasiconformal groups
$(QConf(J))$ associated with non-Euclidean Jordan algebras of degree
three.
}
\end{table}
\end{small}

\section{Quasiconformal Groups associated with  Euclidean Jordan Algebras of Degree Three}
\subsection{Quasiconformal Group Actions and Quartic Light-cones}
\renewcommand{\theequation}{\arabic{section}.{\arabic{subsection}}.\arabic{equation}}
\setcounter{equation}{0}
General theory of novel quasiconformal realizations of Lie groups over
Freudenthal triple systems extended by an extra singlet coordinate was
given in \cite{Gunaydin:2000xr} which we review in appendix C. Since the
automorphism group of a Freudenthal triple system $\mathcal{F}(J)$
defined over a Jordan algebra $J$ of degree three is isomorphic to the
four dimensional U-duality group $G_4$ of MESGT defined by $J$
the original formulation of \cite{Gunaydin:2000xr} is covariant with
respect to  $G_4$.
  In this section we shall study quasiconformal realizations
of groups associated with FTS's defined by Jordan algebras of degree
three in a basis covariant with respect to U-duality groups of five
dimensional $N=2$ MESGTs with symmetric target spaces.  For
convenience we shall label the elements of FTS $\mathcal{F}(J)$
defined over $J$ as follows
\begin{equation}
    \begin{pmatrix}
        \alpha & \mathbf{x} \cr
    \mathbf{y} & \beta
    \end{pmatrix} \equiv \left(\alpha, \beta, \mathbf{x}, \mathbf{y}\right)
\end{equation}
where $\alpha, \beta \in \mathbb{R}$ and $\mathbf{x}, \mathbf{y} \in
J$. The skew-symmetric bilinear form of two elements
$X=\left(\alpha,\beta,\mathbf{x},\mathbf{y}\right)$ and
$Y=\left(\gamma,\delta,\mathbf{w},\mathbf{z}\right)$ of
$\mathcal{F}(J)$ is given by
\begin{equation}
  \left< X, Y\right> \equiv
         \alpha \delta - \beta\gamma +
         \left(\mathbf{x}, \mathbf{z}\right) -
     \left(\mathbf{y}, \mathbf{w}\right)
\end{equation}
where $\left(\mathbf{x}, \mathbf{z}\right)$ is the symmetric bilinear
form over $J$ defined by the trace form $\mathrm{Tr}$ of $J$:
\begin{equation}
    \left( \mathbf{x}, \mathbf{z}\right) \equiv \mathop\mathrm{Tr}\left( \mathbf{x} \circ \mathbf{z}\right)
\end{equation}
The quartic norm $\mathcal{Q}_4 \left(X\right)$ of
$X=\left(\alpha,\beta,\mathbf{x},\mathbf{y}\right)$ is defined as
\begin{equation}
\mathcal{Q}_4 \left(X\right) \equiv \frac{1}{48} \left< \left(X, X, X\right), X \right>
\end{equation}
where $\left(X, Y, Z\right)$ denotes the Freudenthal triple product.
For FTS's defined over Jordan algebras $J$ of degree three the
quartic norm of an element
$X=\left(\alpha,\beta,\mathbf{x},\mathbf{y}\right)$
takes the following form \cite{MR0295205}
\begin{equation}
\mathcal{Q}_4 \left(X\right) = \frac{1}{8} \left\{ \left( \alpha\beta - \left(\mathbf{x}, \mathbf{y}\right)\right)^2 -
        4 \left(\mathbf{x}^\natural, \mathbf{y}^\natural\right) + 4 \alpha \mathbf{N}\left(\mathbf{y}\right)
        + 4 \beta \mathbf{N}\left(\mathbf{x}\right) \right\} \label{Q4}
\end{equation}
where the adjoint map $\natural : J \to J$ is defined such that
\begin{equation}
   \left(\mathbf{x}^\natural\right)^\natural  = \mathbf{N}\left(\mathbf{x}\right) \, \mathbf{x}
\end{equation}
and $\mathbf{N}(\mathbf{x})$ is the cubic norm of the element
$\mathbf{x}\in J$. Using the adjoint map one can define the symmetric
Freudenthal product $\natural$ among two elements of $J$ :
\begin{equation}
  \mathbf{x} ~ \natural  ~ \mathbf{y}  =
      ( \mathbf{x} + \mathbf{y} )^{\natural} - (\mathbf{x})^\natural - (\mathbf{y})^\natural
\end{equation}
For simple Jordan algebras $J_3^{\mathbb{A}} $ of degree three the
$\natural $ product can be expressed in terms of the Jordan product
and trace form $\mathrm{Tr}$ as follows:
\begin{equation}
   \mathbf{x} ~ \natural ~\mathbf{y} = 2 \mathbf{x} \circ \mathbf{y} -
         \mathop\mathrm{Tr}\left(\mathbf{x}\right) \mathbf{y} -
     \mathop\mathrm{Tr}\left(\mathbf{y}\right) \mathbf{x} +
     \left( \mathop\mathrm{Tr}\left(\mathbf{x}\right) \mathop\mathrm{Tr}\left(\mathbf{y}\right) -
       \left( \mathbf{x}, \mathbf{y} \right) \right) \mathop\mathrm{I_3}
\end{equation}
where $\mathrm{I_3}$ is the identity element of $J$. The adjoint
$\mathbf{x}^\natural$ of $\mathbf{x}$ is then
\begin{equation}
    \mathbf{x}^\natural = \frac{1}{2} ( \mathbf{x} ~ \natural ~ \mathbf{x} )
\end{equation}
Referring to Appendix C for a brief summary and to the original
references \cite{Gunaydin:2000xr,Gunaydin:2005zz} for details let us
write down the action of Lie algebra of the quasiconformal group,
associated with a FTS $\mathcal{F}$, on
the vector space $\mathcal{T}$ coordinatized by elements $X$ of
$\mathcal{F}$ and an extra single variable $x$
\cite{Gunaydin:2000xr,Gunaydin:2005zz}:
\begin{equation}
\begin{split}
  \begin{aligned}
      K\left(X\right) &= 0 \\
      K\left(x\right) &= 2\,
  \end{aligned}
  & \quad
  \begin{aligned}
     U_A \left(X\right) &= A \\
     U_A\left(x\right) &= \left< A, X\right>
  \end{aligned}
   \quad
   \begin{aligned}
      S_{AB}\left(X\right) &= \left( A, B, X\right) \\
      S_{AB}\left(x\right) &= 2 \left< A, B\right> x
   \end{aligned}
 \\ \label{qcg}
 &\begin{aligned}
    \Tilde{U}_A\left(X\right) &= \frac{1}{2} \left(X, A, X\right) - A x \\
    \Tilde{U}_A\left(x\right) &= -\frac{1}{6} \left< \left(X, X, X\right), A \right> + \left< X, A\right> x
 \end{aligned}
 \\
 &\begin{aligned}
    \Tilde{K}\left(X\right) &= -\frac{1}{6} \,  \left(X,X,X\right) +  X x \\
    \Tilde{K}\left(x\right) &= \frac{1}{6} \,  \left< \left(X, X, X\right), X \right> + 2\,  \, x^2
 \end{aligned}
\end{split}
\end{equation}
where $A,B \in \mathcal{F}$.
 The  quartic norm over the space
$\mathcal{T}$ is defined as
\begin{equation}
\cN_4(\cX) := \mathcal{Q}_4(X) - x^2
\end{equation}
where $\mathcal{Q}_4(X)$ is the quartic invariant of $\mathcal{F}$.
Quartic ``symplectic distance'' $d(\cX,\cY)$ between any two points
$\cX=(X,x)$ and $\cY=(Y,y) $ in $\mathcal{T}$ is defined as the
quartic norm of ``symplectic difference'' $\gd(\cX,\cY):=
(X-Y,x-y+\langle X, Y \rangle )$ of two vectors
\begin{equation}
  d(\cX,\cY):= \cN_4(\gd(\cX,\cY) = \mathcal{Q}_4 (X-Y) - \left( x-y+\langle X,Y \rangle \right)^2
  \label{quarticlightcone}
\end{equation}
Invariance of this quartic ``symplectic distance function'' $d(\cX,\cY)$
under the action of automorphism group of $\mathcal{F}$ generated by
$S_{(AB)}$ and under ``symplectic translations'' generated by $U_A$ and
$K$ is manifest. Generator $\Delta$ simply rescales $d(\cX,\cY)$ and
under the action of generators $\tilde{U}_A$ and $\tilde{K}$ the
distance function $d(\cX,\cY)$ gets multiplied by terms linear $\cX$
and $\cY$. Hence the quasiconformal group action preserves light-like
separations
\begin{equation*}
  d(\cX,\cY)=0
\end{equation*}
with respect to quartic distance function \cite{Gunaydin:2000xr}. Hence we
shall refer to the quartic distance function \ref{quarticlightcone} as
``quartic light-cone''.  Quasiconformal realization of a simple Lie
algebra $\mathfrak{g}$ over a FTS $\mathcal{F}$ extended by an extra
coordinate $x$ carries over, in a straightforward manner, to that of the
complex Lie algebra $\mathfrak{g}(\mathbb{C})$ by complexifying
$\mathcal{F}$ and $x$.

As discussed above logarithm of cubic distance function $\mathcal{V}(z
- \bar{z})$, written in complex coordinates, that we refer to as
``cubic light cone'',is simply the K\"ahler potential of complex
special geometry of four dimensional MESGT that descend from five
dimensional MESGT uniquely determined by the C-tensor. This holds true
for all four dimensional theories that descend from five dimensions and not
only for those theories with symmetric scalar manifolds
\cite{Gunaydin:1983bi}. Translational symmetries of the cubic
light-cone (hence the K\"ahler potential) follows from Abelian gauge
symmetries of the five dimensional supergravity theories.For theories
with symmetric scalar manifolds defined by Jordan algebras there are
additional symmetries, namely the special conformal transformations
which lead to K\"ahler transformations under which the K\"ahler metric
is invariant. There is a parallel picture in going from four to three
dimensions. In four dimensional MESGT's defined by Freudenthal triple
systems $\mathcal{F}$ the space coordinatized by real coordinates of
$\cX =(X,x)$ on which the three dimensional U-duality group acts as a
quasiconformal group correspond to the boundary coordinates of the
twistor space \cite{Gunaydin:2007qq}
\begin{equation}
\mathcal{Z}_3 = \frac{QConf(J)}{\widetilde{Conf}(J)\times SU(2)} \times \frac{SU(2)}{U(1)} = \frac{{QConf}(J)}{U(1)} \label{twistorspace}
\end{equation}
 of the quaternionic symmetric scalar manifold
\begin{equation*}
 \mathcal{M}_3=  \frac{QConf(J)}{\widetilde{Conf}(J)\times SU(2)}
\end{equation*}
Since the quasiconformal action of a group $G$ extends to its complexification, the complex coordinates of the twistor space can be taken to be  the complex extensions of the coordinates $\cX = (X,x) $ which we shall denote as $\mathbf{\mathcal{Z}}= (\mathbf{Z}, \mathbf{z})$. The K\"ahler potential of the twistor space \ref{twistorspace} is then simply given by  the logarithm of  quartic distance function \ref{quarticlightcone} written in  complex ``quasiconformal coordinates''\cite{Gunaydin:2007qq}:
\begin{equation}
K(\mathbf{\mathcal{Z}},\mathbf{\mathcal{\bar{Z}}} )= \ln
 d\left(\mathbf{\mathcal{Z}}, \mathbf{\mathcal{\bar{Z}}}\right) = \ln
 \left[ \mathcal{Q}_4 (\mathbf{Z} - \mathbf{\bar{Z}} ) + \left(
 \mathbf{z} - \mathbf{\bar{z}} + \langle \mathbf{Z} , \mathbf{\bar{Z}}
 \rangle \right)^2\right] \label{Kahlerpotential}
\end{equation}
The quartic light-cone is manifestly invariant under the Heisenberg
symmetry group corresponding to ``symplectic translations'' generated
by $U_A$ and $K$ in \ref{qcg}.  ``Symplectic special conformal
generators'' $\Tilde{U}_A$ and $\Tilde{K}$ also form an Heisenberg
subalgebra. The global action of symplectic special conformal
transformations on the quartic light-cone $
d\left(\mathbf{\mathcal{Z}}, \mathbf{\mathcal{\bar{Z}}}\right)$
results in overall multiplicative factors which are holomorphic or
anti-holomorphic \cite{Gunaydin:2007qq}
\begin{equation}
 d\left(\mathbf{\mathcal{Z}}, \mathbf{\mathcal{\bar{Z}}}\right)
  \Longrightarrow f(\mathbf{Z},\mathbf{z}) \bar{f}(
  \mathbf{\bar{Z}},\mathbf{\bar{z}} ) d\left(\mathbf{\mathcal{Z}},
  \mathbf{\mathcal{\bar{Z}}}\right)
\end{equation}
This proves that the light-like separations are left invariant under
the full quasiconformal group action. Furthermore the K\'ahler
potential \ref{Kahlerpotential} of the twistor space undergoes
K\"ahler transformations under the quasiconformal group action and
hence leaves the K\"ahler metric invariant. These results obtained
first for quaternionic symmetric spaces \cite{Gunaydin:2007qq} extend
to general quaternionic manifolds that are in the C-map
\cite{Neitzke:2007ke} in complete parallel to the situation in going
from five to four dimensions. In fact, the correspondence established
between harmonic superspace formulation of $N=2$ sigma models coupled
to $N=2$ supergravity and quasiconformal actions of their isometry
groups \cite{Gunaydin:2007vc} suggests that K\"ahler potentials of
quartic light-cone type should exist for all quaternionic manifolds that
couple to four dimensional $N=2$ supergravity and not only those that are in the C-map.

Let us choose a basis $\mathbf{f}_\alpha \in \mathcal{F}(J)$ for the
FTS and let ${d_{\alpha\beta\gamma}}^\delta$ denote the structure
constants of $\mathcal{F}(J)$ defined as follows
\begin{equation}
   \left(\mathbf{f}_\alpha, \mathbf{f}_\beta, \mathbf{f}_\gamma \right) = {d_{\alpha\beta\gamma}}^\delta \mathbf{f}_\delta
\end{equation}
Under the action of its automorphism group elements of a FTS
$\mathcal{F}$ transform in a symplectic representation. Therefore one
can raise and lower their indices with invariant symplectic metric
$\Omega_{\alpha\beta}$ and its inverse:
\begin{equation}
  \begin{aligned}
      V^\alpha &= \Omega^{\alpha\beta} V_\beta \cr
      V_\alpha &= V^\beta \Omega_{\beta\alpha}
  \end{aligned}
    \qquad
   \begin{aligned}
     & \Omega_{\alpha\beta} \Omega^{\beta\gamma}  = - {\delta_\alpha}^\gamma \cr
     & V_\alpha W^\alpha  = - V^\alpha W_\alpha
   \end{aligned}
\end{equation}
Now
\begin{equation}
  \left< \left( \mathbf{f}_\alpha, \mathbf{f}_\beta, \mathbf{f}_\gamma \right), \mathbf{f}_\delta \right> =
    {d_{\alpha\beta\gamma}}^\epsilon \left< \mathbf{f}_\epsilon, \mathbf{f}_\delta \right> =
    {d_{\alpha\beta\gamma}}^\epsilon \Omega_{\epsilon \delta}  = d_{\alpha\beta\gamma\delta}
\end{equation}
where we assumed the normalization $\left<\mathbf{f}_\alpha,
\mathbf{f}_\beta \right> = \Omega_{\alpha\beta}$.  Thus the quartic
norm $\mathcal{Q}_4\left(X\right)$ of an element $X = X^\alpha
\mathbf{f}_\alpha \in \mathcal{F}$ is given as
\begin{equation}
   \mathcal{Q}_4 = \frac{1}{48} \left< \left(X, X,X\right) , X \right> =
         \frac{1}{48} d_{\alpha\beta\gamma\delta} X^\alpha X^\beta X^\gamma X^\delta =
         \frac{1}{48} S_{\alpha\beta\gamma\delta} X^\alpha X^\beta X^\gamma X^\delta
\end{equation}
where $S_{\alpha\beta\gamma\delta} \equiv
d_{\left(\alpha\beta\gamma\delta\right)}$ is the completely
symmetrized structure constants $d_{\alpha\beta\gamma\delta}$ and can
be written as
\begin{equation}
  S_{\alpha\beta\gamma\delta} = d_{\alpha\beta\gamma\delta}  + \text{products of } \Omega \, \Omega
\end{equation}

\subsection{Quasiconformal Lie Algebras twisted by a unitary character}
\renewcommand{\theequation}{\arabic{section}.{\arabic{subsection}}.\arabic{equation}}
\setcounter{equation}{0}
 As already stated above automorphism group $\mathop\mathrm{Aut}(\mathcal{F}(J))$ of a
Freudenthal triple system $\mathcal{F}(J)$ defined over Jordan
algebras $J$ of degree three is isomorphic to the conformal group
$\mathop\mathrm{Conf}(J)$ of $J$:
\begin{equation*}
   \mathrm{Conf}(J) \cong \mathrm{Aut}( \mathcal{F}(J) )
\end{equation*}
 The reduced structure group $\mathop\mathrm{Str}_0\left(J\right)$ is
a subgroup of $\mathop\mathrm{Conf}(J)$ under which the elements $J$
and $\tilde{J}$ transform in conjugate representations. Hence we shall
label the basis vectors of $\mathcal{F}(J)$ as follows
\begin{equation}
 \begin{pmatrix} \alpha & \mathbf{x} \cr \mathbf{y} & \beta \end{pmatrix} = \alpha \mathbf{e}_0 + \beta \tilde{\mathbf{e}}^0
      + x^I \mathbf{e}_I + y_I \tilde{\mathbf{e}}^I
\end{equation}
where $I=1, \dots,  n_V$  and $n_V= dim(J)$.

The quasiconformal Lie algebra $QConf(\mathcal{F}(J))$ can be given a
7 by 5 graded decomposition that is covariant with respect to the
reduced structure group $\mathop\mathrm{Str}_0(J) $ as shown in Table
\ref{7by5grading}.
\begin{table} 
\[
\begin{array}{ccccccc}
 \phantom{U}~~~ & \phantom{R}~~~ & \phantom{U}~~~ &  K ~~~ & \phantom{V}~~~ & \phantom{R}~~~ & \phantom{V}~~~ \\[4pt]
  U_0 & \phantom{R} & U_I & \vline \phantom{K} & V^I & \phantom{R} & V^0 \\[4pt]
 --- &  \Tilde{R}^I  &  --- & \left(  \mathcal{D} \oplus R_I^J  \right) &
             --- & R_J &  --- \\[4pt]
  \Tilde{U}_0 & \phantom{R} & \Tilde{U}_I & \vline \phantom{K} & \Tilde{V}^I & \phantom{R} & \Tilde{V}^0 \\[4pt]
  \phantom{U} & \phantom{R} & \phantom{U} &  \Tilde{K} & \phantom{V} & \phantom{R} & \phantom{V}
\end{array}
\]
\caption{\label{7by5grading}
Above we give the $7 \times 5$ grading of the quasiconformal Lie
algebra associated with the Freudenthal triple system
$\mathcal{F}(J)$. The vertical 5-grading is determined by
$\mathcal{D}=-\Delta$ that commutes with the generators $R_I^J$ of the
structure group of $J$ and the horizontal 7-grading is determined by
$\mathcal{R} \equiv \frac{1}{n_V} {R^I}_I$, which commutes with the
generators of the reduced structure group $Str_0(J)$ of $J$.  The
generators $\Tilde{R}^I, R_I^J $ and $R_I$ generate the automorphism
group of the Freudenthal triple system $\mathcal{F}(J)$ that is
isomorphic to the conformal group $\mathop\mathrm{Conf}(J)$ of the
underlying Jordan algebra $J$, under which the grade +1 (-1)
generators with respect to $\mathcal{D}$, namely $U_0,U_I, V^I, V^0 $
($ \Tilde{U}_0, \Tilde{U}_I , \Tilde{V}^I , \Tilde{V}^0 $) transform
linearly in a symplectic representation.}
\end{table}
With applications to supergravity theories in mind we shall label the
elements $X,Y,..$ of FTS $\mathcal{F}(J)$ in terms of coordinates
($q_0$, $q_I$) and momenta ($p^0$, $p^I$) as follows\footnote{ $ p^I
q_I = \left(\mathbf{e}_I p^I, q_I \Tilde{\mathbf{e}}^I \right)$ and $
\left(p^\sharp\right)_I \left(q^\sharp\right)^I = \left( {p^\sharp}_I
\Tilde{\mathbf{e}}^I, {q^\sharp}^I \mathbf{e}_I \right)$.  }
\begin{equation}
    X = q_0 \Tilde{\mathbf{e}}^0 + q_I \Tilde{\mathbf{e}}^I + p^I \mathbf{e}_I + p^0 \mathbf{e}_0
\end{equation}
We shall normalize the basis elements and cubic norm (C-tensor) such
that the quartic invariant is given by \footnote{ We should note that
$C_{IJK} = C^{IJK}$ is an invariant tensor of
$\mathop\mathrm{Str}_0(J)$.}
\begin{eqnarray}
I_4(X)& = &
    \left(p^0 q_0 -  p^I q_I \right)^2 - \frac{4}{3}  C_{IJK} p^J p^K C^{ILM} q_L q_M \\  \nonumber && +
             \frac{4}{3\sqrt{3}} p^0 C^{IJK} q_I q_J q_K + \frac{4}{3\sqrt{3}} q_0 C_{IJK} p^I p^J p^K \\ \nonumber
             &=& \left(p^0 q_0 -  p^I q_I \right)^2 - \frac{4}{3} (p^\sharp)_I  (q^\sharp)^I \\ && +
             \frac{4}{3\sqrt{3}} p^0 \mathcal{N}(q) + \frac{4}{3\sqrt{3}} q_0 \mathcal{N}(p)
         \nonumber
\end{eqnarray}
where
\begin{equation*}
 \begin{array}{ccc}
   \mathcal{N}(q) \equiv C^{IJK} q_I q_J q_K  & \phantom{seven letters}
   &(q^\sharp)^I   \equiv C^{IJK} q_J q_K \\
   \mathcal{N}(p) \equiv C_{IJK} p^I p^J p^K  &  &(p^\sharp)_I \equiv C_{IJK} p^J p^K
 \end{array}
\end{equation*}
The quartic invariant $I_4$ agrees with the quartic invariant \ref{Q4}
used in mathematics literature and in \cite{Gunaydin:2000xr} up to an
overall factor .  $I_4(X) = 8 \mathcal{Q}_4 \left(X \right)$ under the
identifications
\begin{equation*}
   \mathbf{N}(\mathbf{x}) = \frac{1}{3\sqrt{3}} \mathcal{N}(\mathbf{x})
\end{equation*}
\begin{equation*}
  \mathbf{x}^\natural = \frac{1}{\sqrt{3}} \mathbf{x}^\sharp
\end{equation*}

The normalization of basis vectors $\mathbf{e}_I$
($\Tilde{\mathbf{e}}^I$) of the Jordan algebra $J$ (and its conjugate
$\Tilde{J}$) such that
\begin{eqnarray}
    \left(\Tilde{e}^J, \Tilde{e}^I  \right) =  \mathop\mathrm{Tr} \Tilde{e}^I \circ \Tilde{e}^J =
            \delta^{IJ} \\
  \left( e_I,e_J\right) = \mathop\mathrm{Tr} e_I \circ e_J = \delta_{IJ}
\end{eqnarray}

The action of quasiconformal group $QConf(J)$ on the space
$\mathcal{T} = \mathcal{F}(J) \oplus \mathbb{R}$ with coordinates
$q_0, q_I, p^0, p^I$ of $\mathcal{F}(J)$ plus an extra singlet
coordinate $x \in \mathbb{R}$, twisted by a unitary character $\nu$,
is given by the following differential operators:
\begin{eqnarray}
K&=&\partial_x \\
U_0 &=& \partial_{p^0} + q_0 \partial_x  \\
U_I &=& -  \partial_{p^I} +  q_I \partial_x \\
V^0 &=& \partial_{q_0} - p^0 \partial_x \\
V^I &=&  \partial_{q_I} +  p^I \partial_x \\
   R_I &= & -\sqrt{2} C_{IJK} p^K \partial_{q_K} - \sqrt{\frac{3}{2}}
   \left( p^0 \partial_{p^I} + q_I \partial_{q_0}\right) \\
   \Tilde{R}^I & =& \sqrt{2} C^{IJK} q_J \partial_{p^K} +
   \sqrt{\frac{3}{2}} \left( q_0 \partial_{q_I} + p^I \partial_{p^0}
   \right) \\
   {R_I}^J& = &\frac{3}{2} {\delta_I}^J \left( p^0 \partial_{p^0} -
             q_0 \partial_{q_0}\right) + \\ \nonumber && \frac{3}{2}
             \left( {\delta_I}^N {\delta_K}^J - \frac{4}{3} C_{IKL}
             C^{JNL} \right) \left( q_N \partial_{q_K} - p^K
             \partial_{p^N}\right) \\
  \mathcal{R} &=& \frac{1}{n_V} {R_I}^I = \frac{3}{2} \left( p^0
                       \partial_{p^0} - q_0 \partial_{q_0}\right) +
                       \frac{1}{2} \left( p^I \partial_{p^I} - q_I
                       \partial_{q^I}\right) \\ && \nonumber \\
   \Delta & = &=-\mathcal{D}= -\left( p^0 \partial_{p^0} + p^I
      \partial_{p^I} + q_0 \partial_{q_0} + q_I \partial_{q_I} - \nu
     \right)-2x\partial_{x} \\
  \nonumber && \\
  \Tilde{K} &= & x \left( p^0 \partial_{p^0} + p^I \partial_{p^I} +
               q_0 \partial_{q_0} + q_I \partial_{q_I} - \nu \right) +
               \left( x^2 + I_4 \right) \partial_x \nonumber \\ & & +
               \frac{1}{2} \left( \frac{ \partial{I_4}}{\partial p^0}
               \partial_{q_0} - \frac{ \partial{I_4}}{\partial q_0}
               \partial_{p^0} + \frac{ \partial{I_4}}{\partial q_I}
               \partial_{p^I} - \frac{ \partial{I_4}}{\partial p^I}
               \partial_{q_I} \right)
\end{eqnarray}

The vertical five grading is given by the adjoint action of $\mathcal{D}$
\begin{eqnarray}
\left[ \mathcal{D}, \left(\begin{array}{c} U_0 \\U_I \\V^I  \\ V^0 \end{array}\right) \right] =
 \left(\begin{array}{c} U_0 \\U_I \\V^I \\ V^0 \end{array}\right)
\end{eqnarray}
The grade $-1$ generators (with respect to $\mathcal{D}=-\Delta$ ) are  obtained from the above expressions by commutation with the grade $-2$ generator $\Tilde{K}$
\begin{eqnarray}
    \Tilde{U}_0 &=& \left[ U_0, \Tilde{K}\right] \\
    \Tilde{V}^0 &=& \left[ V^0, \Tilde{K} \right] \\
    \Tilde{U}_I &=& \left[ U_I, \Tilde{K}\right] \\
    \Tilde{V}^I &=& \left[ V^I, \Tilde{K} \right]
\end{eqnarray}
and satisfy
\begin{eqnarray}
[ \mathcal{D}, \left(\begin{array}{c} \Tilde{U}_0 \\ \Tilde{U}_I\\ \Tilde{V}^I   \\ \Tilde{V}^0 \end{array}\right) ] = -\left(\begin{array}{c} \Tilde{U}_0 \\ \Tilde{U}_I  \\ \Tilde{V}^I  \\ \Tilde{V}^0 \end{array}\right)
\end{eqnarray}
The remaining non-vanishing commutation relations of Lie algebra of $QConf(J)$  are as follows:
\begin{eqnarray}
 \left[ K , \Tilde{K} \right] &=& \Delta \\
 \left[ \Delta, K \right] &=& -  2K \\
 \left[ \Delta, \Tilde{K} \right] &=& 2 \Tilde{K} \\
\left[ U_I,V^J \right] &=& -2 \delta_I^J K \\
\left[ U_0, V^0 \right] &=& -2 K \\
\left[ K, \Tilde{U}_0 \right] &=& U_0 \\
\left[ K, \Tilde{U}_I \right] &=& U_I \\
\left[ K, \Tilde{V}^I \right] &=& V^I \\
\left[ K, \Tilde{V}^0 \right] &=& V^0\\
\left[ \Tilde{U}_I , \Tilde{V}^J \right] &=&- 2 \delta_I^J \Tilde{K} \\
\left[ \Tilde{U}_0 , \Tilde{V}^0 \right] &=&- 2  \Tilde{K} \\
    \left[ U_0 ,\Tilde{V}^0 \right] &=& -2 \mathcal{R} + \mathcal{D} \\
   \left[ V^0 ,\Tilde{U}_0 \right] &=& -2 \mathcal{R} - \mathcal{D} 
\end{eqnarray}
\begin{eqnarray}
\left[ R_I^J, R_K \right] &=&
     \frac{3}{2} \Lambda_{IK}^{JL} R_L \\
\left[ R_I^J, \tilde{R}^L \right]& =&
    -\frac{3}{2} \Lambda_{IL}^{JK} \tilde{R}^L \\
\left[ R_I^J, U_K \right] &=&
     \frac{3}{2} \Lambda_{IK}^{JL} U_L -
     \frac{3}{2} \delta_I^J U_K \\
\left[ R_I^J, V^K \right]& =&
     -\frac{3}{2} \Lambda_{IL}^{JK} V^L +
     \frac{3}{2} \delta_I^J V^K \\
\left[ R_I^J, \tilde{U}_K \right] &=&
      \frac{3}{2} \Lambda_{IK}^{JL} \tilde{U}_L -
      \frac{3}{2} \delta_I^J \tilde{U}_K \\
\left[ R_I^J, \tilde{V}^K \right] & =&
     -\frac{3}{2} \Lambda_{IL}^{JK} \tilde{V}^L +
     \frac{3}{2} \delta_I^J \tilde{V}^K \\
\left[ R_I, \tilde{R}^J \right] &=& - R_I^J \\
\left[ U_0, \tilde{V}^I \right] &= &
      2\sqrt{\frac{2}{3}} \tilde{R}^I \\
\left[ \tilde{U}_0, V^I \right] &=&
      -2\sqrt{\frac{2}{3}} \tilde{R}^I  \\
\left[ V^0, \tilde{U}_I \right] &=&
      -2\sqrt{\frac{2}{3}} R_I    \\
\left[ \tilde{V}^0, U_I \right] &=&
      -2\sqrt{\frac{2}{3}} R_I             \\
\left[ U_I, \tilde{V}^J \right] &=&
       \frac{4}{3} R_I^J -
       \delta_I^J \left( \Delta + 2 {\mathcal R} \right) \\
\left[ \tilde{U}_I, V^J \right] &=&
       -\frac{4}{3} R_I^J -
       \delta_I^J \left( \Delta - 2 {\mathcal R} \right)  \\
\left[ U_I, \tilde{U}_J \right] &=&
     - \frac{4}{3} \sqrt{2} C_{IJK} \tilde{R}^K \\
\left[ V^I, \tilde{V}^J \right] &=&
     - \frac{4}{3} \sqrt{2} C^{IJK} R_K   \\
\left[ V^I, R_J \right] &=&
      -\sqrt{\frac{3}{2}} \delta^I_J V^0   \\
\left[ \tilde{V}^I, R_J \right] &=&
      -\sqrt{\frac{3}{2}} \delta^I_J \tilde{V}^0  \\
\left[ \tilde{U}_I, \tilde{R}^J \right] &=&
      -\sqrt{\frac{3}{2}} \delta_I^J \tilde{U}_0  \\
\left[ U_I, \tilde{R}^J \right] &=&
      -\sqrt{\frac{3}{2}} \delta_I^J  U_0 \\
\end{eqnarray}
where
\begin{equation}
   \Lambda_{KL}^{IJ} := \delta_K^I \delta_L^J + \delta_L^I \delta^J_K - \frac{4}{3} C^{IJM} C_{KLM}
\end{equation}
The generators $K, \Tilde{K}, U_0, \Tilde{U}_0, V^0, \Tilde{V}^0,
\mathcal{R}$ and $\Delta$ form the Lie subalgebra $SL(3,\mathbb{R})$
whose maximal compact subalgebra $SO(3)$ is generated by
\begin{eqnarray}
T_1 &:= & \frac{1}{\sqrt{2}} \left( U_0 - \Tilde{V}^0 \right)  \\
T_2 & := & \frac{1}{\sqrt{2}} \left( V^0 + \Tilde{U}_0 \right) \\
T_3 & := & - \left( K + \Tilde{K} \right)
\end{eqnarray}
They satisfy the commutation relations
\begin{equation}
\left[ T_i , T_j \right] =\epsilon_{ijk} T_k
\end{equation}
where $i,j,k=1,2,3$.
The noncompact generators of $SL(3,\mathbb{R})$ are $\mathcal{R}$, $(K
-\Tilde{K})$, $\Delta$, $(U_0 +\Tilde{V}^0)$ and $(V^0 -
\Tilde{U}_0)$. They transform in the spin 2 representation under
$SO(3)$ subgroup. Centralizer of this $SL(3,\mathbb{R})$ subgroup
inside $QConf(J)$ is the reduced structure group $Str_0(J)$ of the
underlying Jordan algebra $J$:
\begin{equation}
QConf(J)\supset Str_0(J) \times SL(3,\mathbb{R})
\end{equation}
Maximal compact subgroups of quasiconformal groups $QConf(J)$
associated with Euclidean Jordan algebras of degree three are of the
form
\begin{equation*}
   \widetilde{Conf(J)}\times SU(2)_L  \subset QConf(J)
\end{equation*}
where the $\widetilde{Conf(J)}$ is the compact real form of the
conformal group of the Jordan algebra $J$.  Hence the quotient
\begin{equation*}
  \frac{QConf(J)}{\widetilde{Conf(J)}\times SU(2)_L}
\end{equation*}
is a quaternionic symmetric space whose coset generators transform in
the $(2n_V + 2,2) $ representation of $ \widetilde{Conf(J)}\times
SU(2)_L$.
The generators of the maximal compact subgroup $\widetilde{Conf}(J)
\times SU(2)_L$ are as follows:
\begin{eqnarray}
&(U_I-\Tilde{V}^I), \nonumber \\ \nonumber
&(V^I + \Tilde{U}_I), \\ \nonumber & (R_I + \Tilde{R}^I ), \\ \nonumber & A_{IJ}=-A_{JI}=(R_I^J-R_J^I), \\
& (\Tilde{U}_0 + V^0 ), \\ \nonumber
& (-U_0 + \Tilde{V}^0), \\ \nonumber
 & (K + \Tilde{K})
\end{eqnarray}
Automorphism group of the underlying Jordan algebra is generated by
$A_{IJ}$ and is compact for Euclidean Jordan algebras. The compact
structure group $\widetilde{Str}(J)$ is a subgroup of $
\widetilde{Conf}(J)$ and is generated by $A_{IJ}$ and $(R_I +
\Tilde{R}^I)$:
\begin{equation}
\widetilde{Str}(J) \Longleftrightarrow A_{IJ} \oplus (R_I +\Tilde{R}^I) \qquad ; I,J=1,..,n_V
\end{equation}
$SU(2)_L$ subalgebra is generated by the following linear combinations
\begin{eqnarray}
L_3 & =&  1/4 \left( K + \Tilde{K} + \sqrt{\frac{2}{3}} \left(\sum_{I=1}^{n_V} ( R_I + \Tilde{R}^I ) \right)\right) \nonumber \\
L_1  &= &\frac{1}{4\sqrt{2}} \left(-U_0 + \Tilde{V}^0 + \sum_{I=1}^{n_V} ( \Tilde{U}_I + V^I ) \right) \\
L_2 &= & \frac{1}{4\sqrt{2}} \left( \Tilde{U}_0 + V^0 + \sum_{I=1}^{n_V} ( U_I -\Tilde{V}^I ) \right) \nonumber
\end{eqnarray}
and satisfy the commutation relations
\begin{equation}
  \left[L_i, L_j \right] = \epsilon_{ijk} L_k \qquad ; i,j,\ldots=1,2,3
\end{equation}
 We will label the basis elements of the Jordan algebra $J$ such that  $e_1, e_2$ and $e_3$ are the three irreducible  idempotents of  $J$ and the identity element $\mathbb{I}$   is simply
\begin{equation}
  \mathbb{I} = e_1 + e_2 + e_3
\end{equation}
With this labelling  the operators
\begin{equation*}
  \sum_{i=1}^{3} (U_i-\Tilde{V}^i) \;\; , \;\; \sum_{i=1}^3 (V^i+\Tilde{U}_i) \;\;\; , \;\; \sum_{i=1}^3(R_i+\Tilde{R}^i)
\end{equation*}
generate an $SU(2)_S$ subgroup of $\widetilde{Conf(J)}$ whose
centralizer inside $\widetilde{Conf}(J)$ is the automorphism group
$Aut(J)$ of Jordan algebra $J$ generated by $A_{IJ}=R_I^J-R_J^I$
Thus we have the inclusions
\begin{equation}
  QConf(J) \supset \widetilde{Conf(J)}\times SU(2)_L \supset Aut(J) \times SU(2)_S \times SU(2)_L
\end{equation}
The centralizer of $Aut(J)$ inside the full  $QConf(J)$ is the split exceptional group $G_{2(2)}$:
\begin{equation}
  QConf(J)\supset Aut(J) \times G_{2(2)}
\end{equation}
\subsection{Quadratic Casimir Operators of Quasiconformal Lie algebras}
\renewcommand{\theequation}{\arabic{section}.{\arabic{subsection}}.\arabic{equation}}
\setcounter{equation}{0}
The generators $S_I^J$ of the reduced structure ( Lorentz) group of a Jordan algebra are given by the traceless components of $R_I^J$:
\begin{equation}
S_I^J = R_I^J - \frac{1}{n_V} \delta_I^J ( R_K^K) = R_I^J - \delta_I^J \mathcal{R}
\end{equation}
The quadratic Casimir operator of the quasiconformal group $QConf(J)$ of a simple Jordan algebra $J$ of degree three can then be written in a general form involving a single parameter $\alpha$ :

\begin{eqnarray}
\mathcal{C}_2 &= &\alpha S_I^J S_J^I - \frac{4}{3} ( \mathcal{R}^2 + \Tilde{R}^I R_I + R_I \Tilde{R}^I )
+ ( U_0 \Tilde{V}^0 + U_I \Tilde{V}^I + \Tilde{V}^0 U_0 + \Tilde{V}^I U_I ) \\ \nonumber &&  - (\Tilde{U}_0 V^0 + \Tilde{U}_I V^I + V^0 \Tilde{U}_0 + V^I \Tilde{U}_I )
- 2 ( K \Tilde{K} + \Tilde{K} K) + \Delta^2 
\end{eqnarray}
where $\alpha$ takes on the following  values for different quasiconformal Lie groups $QCG(J)$:
\begin{eqnarray}
\alpha(F_{4(4)})= \frac{16}{45} \\ \nonumber
\alpha(E_{6(2)})= \frac{8}{27} \\ \nonumber
\alpha(E_{7(-5)})= \frac{2}{9} \\ \nonumber
\alpha(E_{8(-24)})= \frac{4}{27} \\ \nonumber
\end{eqnarray}
As for the quasiconformal groups associated with the generic  family of reducible Jordan algebras $J= \mathbb{R} \oplus \Gamma_{(1,n_V-1)}$ 
the quadratic Casimir can be written in the form:
\begin{eqnarray}
\mathcal{C}_2 ( SO(n_V+1,4))  &= &\frac{4}{9} R_I^J R_J^I - \frac{4}{3} ( \Tilde{R}^I R_I + R_I \Tilde{R}^I ) - \frac{(n_V-3)}{9}  (R_2^2 +R_3^3)^2 \nonumber \\ &&
+ ( U_0 \Tilde{V}^0 + U_I \Tilde{V}^I + \Tilde{V}^0 U_0 + \Tilde{V}^I U_I ) \\ \nonumber &&  - (\Tilde{U}_0 V^0 + \Tilde{U}_I V^I + V^0 \Tilde{U}_0 + V^I \Tilde{U}_I )
- 2 ( K \Tilde{K} + \Tilde{K} K) + \Delta^2 
\end{eqnarray}

The quadratic Casimir operators for the quasiconformal realizations given above reduce to a c-number whose value can be expressed universally as 
\eq
\mathcal{C}_2 ( QCG(J)) = \nu (\nu + 2n_V + 4 ) 
\en
where $n_V$ is equal to the dimension $ Dim(J)$ of the underlying Jordan algebra. 
As a consequence the representations induced by the quasiconformal action of $QConf(J)$ over the space of square integrable functions $L^2(p^I,q_I,x) $ are  unitary representations belonging to the principle series under the scalar product
\eq
\langle f|g\rangle = \int \bar{f}(p,q,x) g(p,q,x) dp dq dx 
\en
for 
\eq
\nu = -(n_V+2) + i \rho 
\en
where $\rho \in \mathbb{R}$. For special real discrete values of the twisting parameter $\nu$ one can realize the unitary representations of $QConf(J)$ belonging to the quaternionic discrete series and their continuations over the space of  holomorphic functions of complexified quasiconformal  coordinates. These holomorphic coordinates can be identified with the natural complex coordinates of the twistor spaces associated with the quaternionic symmetric spaces
\[\frac{ QConf(J)}{ \widetilde{Conf}(J) \times SU(2)} \] 
For rank two quaternionic groups these representations were studied in \cite{Gunaydin:2007qq}. Detailed studies of  discrete series representations induced by the quasiconformal actions for higher rank 
groups $QConf(J)$ will be subjects of separate studies.

\section{Jordan Algebra of STU Model and its Quasiconformal Group $SO(4,4)$}
\renewcommand{\theequation}{\arabic{section}.{\arabic{subsection}}.\arabic{equation}}
\setcounter{equation}{0}
To illustrate the general structure of quasiconformal group actions
formulated above we shall study in detail the quasiconformal group
associated with the smallest nontrivial Euclidean Jordan algebra of
degree three 
\begin{equation}
  J= \mathbb{R}\oplus \Gamma_{(1,1)}
\end{equation}
which belongs to the generic Jordan family of reducible Jordan
algebras. Its cubic norm form
\begin{equation}
  \mathcal{N} =\alpha Q(\beta^0,\beta^1)=\alpha [(\beta^0)^2 -(\beta^1)^2]
\end{equation}
can be factorized as
\begin{equation}  \mathcal{N}= \alpha \beta_+ \beta_-
\end{equation}
where $\beta_{\pm}= \beta^0 \pm \beta^1 $.  We shall choose the
normalization of the cubic norm such that the symmetric C-tensor is
given by the absolute value of the Levi-Civita tensor:
\begin{equation}
  C_{IJK} = \frac{\sqrt{3}}{2}| \epsilon_{IJK}|
\end{equation}
so that
\begin{equation}
  C_{IJK} x^Ix^Jx^K = 3\sqrt{3} x^1x^2x^3
\end{equation}
The structure group of the Jordan algebra is
\begin{equation}
  Str(J= \mathbb{R}\oplus \Gamma_{(1,1)})= SO(1,1)\times SO(1,1) \times SO(1,1)
\end{equation}
and the $d=5$, $N=2$ MESGT defined by $J= \mathbb{R}\oplus
\Gamma_{(1,1)}$ describes the coupling of two vector multiplets to
$N=2$ supergravity.  In four dimensions it gives the STU model whose
U-duality group is isomorphic to the the conformal group of $ J=
\mathbb{R}\oplus \Gamma_{(1,1)}$
\begin{equation}
  Conf(J= \mathbb{R}\oplus \Gamma_{(1,1)})= SL(2,\mathbb{R}) \times SL(2,\mathbb{R})
   \times SL(2,\mathbb{R})
\end{equation}
U-duality group of the STU model in three dimensions is  the
quasiconformal group defined over the FTS associated with $J=
\mathbb{R}\oplus \Gamma_{(1,1)}$
\begin{equation}
  QConf(\mathbb{R}\oplus \Gamma_{(1,1)}) = SO(4,4)
\end{equation}

\subsection{Noncompact Basis of Quasiconformal $SO(4,4)$}
In the standard $ 7 \times 5 $-grading of the Lie algebra of $SO(4,4)$ as a quasiconformal Lie algebra
given in Table \ref{SO(4,4)}
\begin{table}
\begin{equation} \nonumber
\begin{array}{ccccccc}
 \phantom{U}~~~ & ~~~\phantom{R}~~~ & \phantom{U}~~~ & K & \phantom{V}~~~ & ~~~\phantom{R}~~~ & ~~~\phantom{V} \\[4pt]
  U_0 & \phantom{R} & U_I & \vline\phantom{K} & V^I & \phantom{R} & V^0 \\[4pt]
 --- \phantom{U} & \Tilde{R}^I &--- \phantom{U} & \left( (SO(1,1))^3 \oplus SO(1,1) \right) &---
             \phantom{V} & R_J &--- \phantom{V}  \\[4pt]
  \Tilde{U}_0 & \phantom{R} & \Tilde{U}_I & \vline \phantom{K} & \Tilde{V}^I & \phantom{R} & \Tilde{V}^0 \\[4pt]
  \phantom{U} & \phantom{R} & \phantom{U} & \Tilde{K} & \phantom{V} & \phantom{R} & \phantom{V}
\end{array}
\end{equation}
\caption{ \label{SO(4,4)} Above we give the 7 by 5 grading of $SO(4,4)$ with respect to the generators $ \mathcal{R}$ and $\mathcal{D}$ respectively. } 
\end{table}
the indices $I,J,..$ run over 1,2,3 and $R_I^J =0$ if $I\neq J$.
Explicit expressions for the generators of grade zero subspace (with
respect to $SO(1,1)$ generator $\mathcal{D} = - \Delta$ ) are as
follows:
\begin{eqnarray}
R_I^I &=& \frac{3}{2} \left(p^0 \frac{\partial}{\partial{p^0}}+ p^K \frac{\partial}{\partial{p^K}}- q_0 \frac{\partial}{\partial{q_0}} - q_K \frac{\partial}{\partial{q_K}} - 2p^I\frac{\partial}{\partial{p^I}} + 2 q_I\frac{\partial}{\partial{q_I}} \right) \\ \nonumber
&& \\ \nonumber && (I=1,2,3) \;\;\;, \text{ I not summed over} \nn
\end{eqnarray}
\begin{equation}
  R_1 = -\sqrt{\frac{3}{2}} \left( p^0 \frac{\partial}{\partial{p^1}}
      +q_1\frac{\partial}{\partial{q_0}} + p^2
      \frac{\partial}{\partial{q_3}} + p^3 \frac{\partial}{\partial{q_2}}
      \right)
\end{equation}
\begin{equation}
    \Tilde{R}^1 = \sqrt{\frac{3}{2}} \left( p^1
       \frac{\partial}{\partial{p^0}} +q_0\frac{\partial}{\partial{q_1}} +
       q_2 \frac{\partial}{\partial{p^3}} + q_3 \frac{\partial}{\partial{p^2}} \right)
\end{equation}
\begin{equation}
      R_2 = -\sqrt{\frac{3}{2}} \left( p^0 \frac{\partial}{\partial{p^2}} +q_2\frac{\partial}{\partial{q_0}} +
      p^3 \frac{\partial}{\partial{q_1}} + p^1 \frac{\partial}{\partial{q_3}}
      \right)
\end{equation}
\begin{equation}
      \Tilde{R}^2 = \sqrt{\frac{3}{2}} \left( p^2 \frac{\partial}{\partial{p^0}} +q_0\frac{\partial}{\partial{q_2}} +
      q_3 \frac{\partial}{\partial{p^1}} + q_1 \frac{\partial}{\partial{p^3}}
      \right)
\end{equation}
\begin{equation}
       R_3 = -\sqrt{\frac{3}{2}} \left( p^0 \frac{\partial}{\partial{p^3}} +q_3\frac{\partial}{\partial{q_0}} +
       p^1 \frac{\partial}{\partial{q_2}} + p^2 \frac{\partial}{\partial{q_1}} \right)
\end{equation}
\begin{equation}
        \Tilde{R}^3 = \sqrt{\frac{3}{2}} \left( p^3 \frac{\partial}{\partial{p^0}} +q_0\frac{\partial}{\partial{q_3}} +
	q_1 \frac{\partial}{\partial{p^2}} + q_2 \frac{\partial}{\partial{p^1}} \right)
\end{equation}
\begin{equation}
        \mathcal{D}= - \Delta = - \left(p^0 \frac{\partial}{\partial{p^0}}+ p^I \frac{\partial}{\partial{p^I}} +
       q_I\frac{\partial}{\partial{q_I}}+ q_0\frac{\partial}{\partial{q_0}}+ 2x \frac{\partial}{\partial{x}} -\nu \right)
\end{equation}
They generate the subgroup $Conf(J)\times SO(1,1) $. The generators of three
$SL(2,\mathbb{R})$ factors of $Conf(J)$ are
\begin{equation}
      SL(2,\mathbb{R})_I \Longleftrightarrow  (R_I \oplus R_I^I \oplus  \Tilde{R}^I ) \;\; , \;\; I=1,2,3
\end{equation}
They satisfy the commutation relations:
\begin{eqnarray}
  \left[R_I,\Tilde{R}^I \right] &=& R_I^I    \\
   \left[ R_I^I , R_I \right] & = & 3 R_I  \\
   \left[ R_I^I,\Tilde{R}^I \right] & = & - 3 \Tilde{R}^I \\
   && \text{I not summed over} \nn
\end{eqnarray}
The linear combinations $(R_I + \Tilde{R}^I)$ for $I=1,2,3$ generate
the compact $U(1)_I$ subgroups of the three $SL(2,\mathbb{R})_I$.
The grade $\pm 2$ generators $K$ and $\Tilde{K}$ together with
$\Delta = - \mathcal{D}$ generate another $SL(2,\mathbb{R})$ subgroup of $SO(4,4)$:
\begin{equation}
     SL(2,\mathbb{R})_0 \longrightarrow K \oplus \Delta \oplus \Tilde{K}
\end{equation}
whose compact $U(1)$ subgroup is generated  by $(K +\Tilde{K})$.
Explicitly we have
\begin{equation}
   K=\frac{\partial}{\partial{x}}
\end{equation}
\begin{equation*}
 \begin{split}
  \Tilde{K} &= x \left( p^0 \partial_{p^0} + p^I \partial_{p^I} + q_0 \partial_{q_0} + q_I \partial_{q_I} - \nu \right)
               + \left( x^2 + I_4 \right) \partial_x \cr
              & + \frac{1}{2} \left(
                    \frac{ \partial{I_4}}{\partial p^0} \partial_{q_0} -
                    \frac{ \partial{I_4}}{\partial q_0} \partial_{p^0} +
                    \frac{ \partial{I_4}}{\partial q_I} \partial_{p^I} -
            \frac{ \partial{I_4}}{\partial p^I} \partial_{q_I}
                     \right)
 \end{split}
\end{equation*}
where the quartic invariant $I_4$ is
\begin{eqnarray}
          I_4 & = 4 p^0 q_1 q_2 q_3 + 4 q_0 p^1 p^2 p^3 + \left(p^0 q_0-p^1 q_1-p^2 q_2-p^3 q_3\right)^2 \\ \nonumber &
                  -4 \left( p^2 q_2 p^3 q_3 +p^1 q_1 p^2 q_2 + p^1 q_1 p^3 q_3\right)
\end{eqnarray}
They satisfy the commutation relations;
\begin{eqnarray}
   \left[ K , \Tilde{K} \right] &=& \Delta \\
   \left[ \Delta, K \right] &=& -  2K \\
   \left[ \Delta, \Tilde{K} \right] &=& 2 \Tilde{K}
\end{eqnarray}
Four mutually commuting $SL(2,\mathbb{R})$ subgroups correspond to the decomposition
\begin{equation}
 SO(4,4)\supset SO(2,2)\times SO(2,2) \cong SL(2,\mathbb{R}) \times
   SL(2,\mathbb{R}) \times SL(2,\mathbb{R}) \times SL(2,\mathbb{R})
\end{equation}
Vertical grade +1 generators (with respect to $\mathcal{D}$ are
\begin{eqnarray}
   U_0 &=& \frac{\partial}{\partial{p^0}}+ q_0 \frac{\partial}{\partial{x}} \\
   V^0 &=& \frac{\partial}{\partial{q_0}} - p^0 \frac{\partial}{\partial{x}}   \\
   U_I &=& -\frac{\partial}{\partial{p^I}}+ q_I \frac{\partial}{\partial{x}}  \\
   V^I &=& -\frac{\partial}{\partial{q_I}}+ p^I \frac{\partial}{\partial{x}}
\end{eqnarray}
The generators belonging to vertical grade $-1$ subspace
 are given  by the differential operators
\begin{eqnarray}
\Tilde{U}_0 &=&  2 q_0^2\left(\frac{\partial \text{}}{\partial q_0}\right) -\nu  q_0+2 q_1 q_0
   \left(\frac{\partial \text{}}{\partial q_1}\right) +2 q_2 q_0
   \left(\frac{\partial \text{}}{\partial q_2}\right) +2 q_3 q_0
   \left(\frac{\partial \text{}}{\partial q_3}\right)  \nonumber \\ && +(-p^0
   q_0+p^1 q_1+p^2 q_2+p^3 q_3+x) \left(\frac{\partial
   \text{}}{\partial p^0}\right) \nonumber \\ && +2 q_2 q_3 \left(\frac{\partial
   \text{}}{\partial p^1}\right)+2 q_1 q_3 \left(\frac{\partial
   \text{}}{\partial p^2}\right)+2 q_1 q_2 \left(\frac{\partial
   \text{}}{\partial p^3}\right) \nonumber \\ && +\left(p^0 q_0^2-p^1 q_1
   q_0-p^2 q_2 q_0-p^3 q_3 q_0+x q_0+2 q_1 q_2
   q_3\right) \left(\frac{\partial \text{}}{\partial x}\right)  \\
\Tilde{V}^0 &=&  -2 (p^0)^2\left(\frac{\partial \text{}}{\partial p^0}\right) +\nu  p^0-2 p^1 p^0
   \left(\frac{\partial \text{}}{\partial p^1}\right) -2 p^2 p^0
   \left(\frac{\partial \text{}}{\partial p^2}\right) -2 p^3 p^0
   \left(\frac{\partial \text{}}{\partial p^3}\right)  \nonumber \\ && +(p^0
   q_0-p^1 q_1-p^2 q_2-p^3 q_3+x) \left(\frac{\partial
   \text{}}{\partial q_0}\right) \nonumber \\ && -2 p^2 p^3 \left(\frac{\partial
   \text{}}{\partial q_1}\right)-2 p^1 p^3 \left(\frac{\partial
   \text{}}{\partial q_2}\right)-2 p^1 p^2 \left(\frac{\partial
   \text{}}{\partial q_3}\right) \nonumber \\ && +(p^1 (2 p^2 p^3-p^0
   q_1)+p^0 (p^0 q_0-p^2 q_2-p^3 q_3-x))
   \left(\frac{\partial \text{}}{\partial x}\right) \\
\Tilde{U}_1 &=&  2 q_1^2\left(\frac{\partial \text{}}{\partial q_1}\right) -\nu  q_1+2 p^2 q_1
   \left(\frac{\partial \text{}}{\partial p^2}\right) +2 p^3 q_1
   \left(\frac{\partial \text{}}{\partial p^3}\right) +2 q_0 q_1
   \left(\frac{\partial \text{}}{\partial q_0}\right) +2 p^2 p^3
   \left(\frac{\partial \text{}}{\partial p^0}\right)\nonumber  \\ && +(p^0 q_0-p^1
   q_1+p^2 q_2+p^3 q_3-x) \left(\frac{\partial \text{}}{\partial
   p^1}\right) \nonumber  +2 p^3 q_0 \left(\frac{\partial \text{}}{\partial
   q_2}\right)+2 p^2 q_0 \left(\frac{\partial \text{}}{\partial
   q_3}\right) \nonumber \\ && +(p^2 (q_1 q_2-2 p^3 q_0)+q_1 (p^0
   q_0-p^1 q_1+p^3 q_3+x)) \left(\frac{\partial \text{}}{\partial
   x}\right)  \\
\Tilde{V}^1 &=&2 (p^1)^2\left(\frac{\partial \text{}}{\partial p^1}\right) -\nu  p^1+2 p^0 p^1
   \left(\frac{\partial \text{}}{\partial p^0}\right) +2 q_2 p^1
   \left(\frac{\partial \text{}}{\partial q_2}\right) +2 q_3 p^1
   \left(\frac{\partial \text{}}{\partial q_3}\right) +2 p^0 q_3
   \left(\frac{\partial \text{}}{\partial p^2}\right) \nonumber \\ && +2 p^0 q_2
   \left(\frac{\partial \text{}}{\partial p^3}\right)+2 q_2 q_3
   \left(\frac{\partial \text{}}{\partial q_0}\right)+(p^0 q_0-p^1
   q_1+p^2 q_2+p^3 q_3+x) \left(\frac{\partial \text{}}{\partial
   q_1}\right) \nonumber \\ && +(p^0 (2 q_2 q_3-p^1 q_0)+p^1 (p^1
   q_1-p^2 q_2-p^3 q_3+x)) \left(\frac{\partial \text{}}{\partial
   x}\right) \\
 \Tilde{U}_2 &=& 2 q_2^2\left(\frac{\partial \text{}}{\partial q_2}\right) -\nu  q_2+2 p^1 q_2
   \left(\frac{\partial \text{}}{\partial p^1}\right) +2 p^3 q_2
   \left(\frac{\partial \text{}}{\partial p^3}\right) +2 q_0 q_2
   \left(\frac{\partial \text{}}{\partial q_0}\right) +2 p^1 p^3
   \left(\frac{\partial \text{}}{\partial p^0}\right) \nonumber \\ && +(p^0 q_0+p^1
   q_1-p^2 q_2+p^3 q_3-x) \left(\frac{\partial \text{}}{\partial
   p^2}\right)+2 p^3 q_0 \left(\frac{\partial \text{}}{\partial
   q_1}\right)+2 p^1 q_0 \left(\frac{\partial \text{}}{\partial
   q_3}\right) \nonumber  \\ && +(p^1 (q_1 q_2-2 p^3 q_0)+q_2 (p^0
   q_0-p^2 q_2+p^3 q_3+x)) \left(\frac{\partial \text{}}{\partial
   x}\right)  \\
   \Tilde{V}^2 &=& 2 (p^2)^2 \left(\frac{\partial \text{}}{\partial p^2}\right) -\nu  p^2+2 p^0 p^2
   \left(\frac{\partial \text{}}{\partial p^0}\right) +2 q_1 p^2
   \left(\frac{\partial \text{}}{\partial q_1}\right) +2 q_3 p^2
   \left(\frac{\partial \text{}}{\partial q_3}\right) +2 p^0 q_3
   \left(\frac{\partial \text{}}{\partial p^1}\right) \nonumber \\ && +2 p^0 q_1
   \left(\frac{\partial \text{}}{\partial p^3}\right)+2 q_1 q_3
   \left(\frac{\partial \text{}}{\partial q_0}\right)+(p^0 q_0+p^1
   q_1-p^2 q_2+p^3 q_3+x) \left(\frac{\partial \text{}}{\partial
   q_2}\right) \nonumber \\ && +(p^0 (2 q_1 q_3-p^2 q_0)+p^2 (-p^1
   q_1+p^2 q_2-p^3 q_3+x)) \left(\frac{\partial \text{}}{\partial
   x}\right)  
   \end{eqnarray}
   \begin{eqnarray} 
\Tilde{U}_3 &=&    2 q_3^2 \left(\frac{\partial \text{}}{\partial q_3}\right) -\nu  q_3+2 p^1 q_3
   \left(\frac{\partial \text{}}{\partial p^1}\right) +2 p^2 q_3
   \left(\frac{\partial \text{}}{\partial p^2}\right) +2 q_0 q_3
   \left(\frac{\partial \text{}}{\partial q_0}\right) +2 p^1 p^2
   \left(\frac{\partial \text{}}{\partial p^0}\right) \nonumber  \\ && +(p^0 q_0+p^1
   q_1+p^2 q_2-p^3 q_3-x) \left(\frac{\partial \text{}}{\partial
   p^3}\right)+2 p^2 q_0 \left(\frac{\partial \text{}}{\partial
   q_1}\right)+2 p^1 q_0 \left(\frac{\partial \text{}}{\partial
   q_2}\right)\nonumber \\ && +(p^1 (q_1 q_3-2 p^2 q_0)+q_3 (p^0
   q_0+p^2 q_2-p^3 q_3+x)) \left(\frac{\partial \text{}}{\partial
   x}\right)  \\
\Tilde{V}^3 &=&  2 (p^3)^2 \left(\frac{\partial \text{}}{\partial p^3}\right) -\nu  p^3+2 p^0 p^3
   \left(\frac{\partial \text{}}{\partial p^0}\right) +2 q_1 p^3
   \left(\frac{\partial \text{}}{\partial q_1}\right) +2 q_2 p^3
   \left(\frac{\partial \text{}}{\partial q_2}\right) +2 p^0 q_2
   \left(\frac{\partial \text{}}{\partial p^1}\right) \nonumber \\ && +2 p^0 q_1
   \left(\frac{\partial \text{}}{\partial p^2}\right)+2 q_1 q_2
   \left(\frac{\partial \text{}}{\partial q_0}\right)+(p^0 q_0+p^1
   q_1+p^2 q_2-p^3 q_3+x) \left(\frac{\partial \text{}}{\partial
   q_3}\right) \nonumber \\ && +(p^0 (2 q_1 q_2-p^3 q_0)+p^3 (-p^1
   q_1-p^2 q_2+p^3 q_3+x)) \left(\frac{\partial \text{}}{\partial
x}\right)
 \end{eqnarray}
The horizontal 7-grading of the Lie algebra of $SO(4,4)$ is given by the adjoint action of
$\mathcal{R}= \frac{1}{3} R^I_I $:
\begin{eqnarray}
   \left[\mathcal{R}, \left(\begin{array}{c} V^0 \\ \Tilde{V}^0 \end{array}\right)\right] & =&
      \frac{3}{2} \left(\begin{array}{c} V^0 \\ \Tilde{V}^0 \end{array}\right)
\end{eqnarray}
\begin{equation}
   \left[ \mathcal{R}, R_I \right]  =  R_I
\end{equation}
\begin{eqnarray}
   \left[ \mathcal{R}, \left(\begin{array}{c} V^I \\ \Tilde{V}^I \end{array}\right) \right] & =&
      \frac{1}{2} \left(\begin{array}{c} V^I \\ \Tilde{V}^I \end{array}\right)
\end{eqnarray}
\begin{equation}
    \left[ \mathcal{R}, R^J_I \right] = 0
\end{equation}
\begin{eqnarray}
   \left[\mathcal{R}, \left(\begin{array}{c} U_I \\ \Tilde{U}_I \end{array}\right)\right] & =&
     -\frac{1}{2} \left(\begin{array}{c} U_I \\ \Tilde{U}_I \end{array}\right)
\end{eqnarray}
\begin{equation}
   \left[\mathcal{R}, \Tilde{R}^I \right] = - \Tilde{R}^I
\end{equation}
\begin{eqnarray}
     \left[\mathcal{R}, \left(\begin{array}{c} U_0 \\ \Tilde{U}_0 \end{array}\right) \right]  &=&
     -\frac{3}{2}\left(\begin{array}{c} U_0 \\ \Tilde{U}_0 \end{array}\right)
\end{eqnarray}
The remaining  nonvanishing commutation relations of $SO(4,4)$ are
\begin{equation}
    \left[ U_0 ,\Tilde{V}^0 \right] = -2 \mathcal{R} + \mathcal{D}
\end{equation}
\begin{equation}
   \left[ V^0 ,\Tilde{U}_0 \right] = -2 \mathcal{R} - \mathcal{D}
\end{equation}
\begin{equation}
    \left[ U_0 , V^0 \right] = - 2 K
\end{equation}
\begin{equation}
   \left[\Tilde{U}_0, \Tilde{V}^0 \right] = - 2 \Tilde{K}
\end{equation}
\begin{equation}
    \left[U_I , \Tilde{V}^J \right] =  \frac{4}{3} R^J_I -\delta^J_I ( 2\mathcal{R} -\mathcal{D})
\end{equation}
\begin{equation}
  \left[ V^I , \Tilde{U}_J \right] = \frac{4}{3} R^I_J -\delta^I_J ( 2 \mathcal{R} + \mathcal{D})
\end{equation}
\begin{equation}
  \left[U_I , \Tilde{U}_J \right] = -2 \sqrt{\frac{2}{3}} \left|\epsilon_{IJK}\right| \Tilde{R}^K
\end{equation}
\begin{equation}
    \left[V^I , \Tilde{V}^J \right] = -2 \sqrt{\frac{2}{3}} \left|\epsilon^{IJK} \right| R_K
\end{equation}
\begin{equation}
    \left[U_0 , \Tilde{V}^I \right] =2 \sqrt{\frac{2}{3}} \Tilde{R}^I
\end{equation}
\begin{equation}
    \left[\Tilde{U}_0, V^I \right] = - 2 \sqrt{\frac{2}{3}} \Tilde{R}^I
\end{equation}
\begin{equation}
    \left[V^0 , \Tilde{U}_I \right] = - 2 \sqrt{\frac{2}{3}} R_I
\end{equation}
\begin{equation}
   \left[ \Tilde{V}^0 , U_I \right] = 2 \sqrt{\frac{2}{3}} R_I
\end{equation}
\begin{eqnarray}
   &[ U_I , R_J ] = \sqrt{\frac{3}{2}} |\epsilon_{IJK}| V^K \\
   &[\Tilde{U}_I , R_J ] =  \sqrt{\frac{3}{2}} |\epsilon_{IJK}|\Tilde{V}^K \\
   &[ V^I , \Tilde{R}^J ] = -\sqrt{\frac{3}{2}} |\epsilon^{IJK}| U_K \\
   &[\Tilde{V}^I , \Tilde{R}^J ]  = -\sqrt{\frac{3}{2}} |\epsilon^{IJK}| \Tilde{U}_K \\
\end{eqnarray}
Recalling that
\begin{equation*}
  \left|\epsilon_{IJK} \right| = \frac{2}{\sqrt{3}} C_{IJK}
\end{equation*}
covariance of the commutation relations above with respect to the
U-duality group $SO(1,1) \times SO(1,1)$ of the five dimensional STU
supergravity model
\cite{Gunaydin:1983bi,Gunaydin:1984ak} becomes manifest. The generators of the maximal
compact subgroup $SO(4)\times SO(4) = SU(2)\times SU(2) \times SU(2)
\times SU(2)$ of $SO(4,4)$ are the following:
\begin{eqnarray}
 &(K+\Tilde{K}), (U_0 -\Tilde{V}^0), (\Tilde{U}_0 + V^0) \nn \\
&(R_I+\Tilde{R}^I),  \nn \\
&(U_I -\Tilde{V}^I),  \nn \\
&(\Tilde{U}_I + V^I ), \nn \\
 & (I=1,2,3)
\end{eqnarray}
Note that the maximal compact subgroup is of the form
\begin{equation*}
  \widetilde{Conf}(J) \times SU(2)_L
\end{equation*}
where $\widetilde{Conf}(J)$ is the compact real form of the conformal
group of the underlying Jordan algebra $J= (\mathbb{R} \oplus
\Gamma_{(1,1)})$
\begin{equation}
\widetilde{Conf}(\mathbb{R} \oplus \Gamma_{(1,1)})= SU(2)_M\times SU(2)_N \times SU(2)_O
\end{equation}
The group $SU(2)_S$ defined in the previous section is simply the
diagonal subgroup of the three $SU(2)$ subgroups and commutes with
$SU(2)_L$.

\subsection{Compact Basis of $SO(4,4)$}
\setcounter{equation}{0}
The generators of the maximal compact subgroup $\widetilde{Conf}
\times SU(2)_L = SU(2)_M\times SU(2)_N \times SU(2)_O \times SU(2)_L $
are given by the following linear combinations of the generators
studied in the previous subsection
\begin{eqnarray}
   & M_3 &:= 1/4 \left(K + \Tilde{K} - \sqrt{2/3} (R_1 + \Tilde{R}^1 + R_2 +
                 \Tilde{R}^2 - R_3 - \Tilde{R}^3)\right) \\
   &M_1 & := \frac{1}{ 4 \sqrt{2}} \left(-U_0 - \Tilde{U}_1  - \Tilde{U}_2 +
                 \Tilde{U}_3 - V^1 - V^2 + V^3 + \Tilde{V}^0 \right) \\
   &M_2 &:= \frac{1}{ 4 \sqrt{2}} \left(-U_1 - U_2  +U_3  + \Tilde{U}_0 +
		 \Tilde{V}^1 +\Tilde{V}^2 - \Tilde{V}^3 + V^0 \right) \\
   &N_3& := 1/4 \left(K + \Tilde{K} - \sqrt{2/3} (R_1 + \Tilde{R}^1 - R_2 -
		 \Tilde{R}^2 + R_3 + \Tilde{R}^3)\right) \\
   &N_1 &:= \frac{1}{ 4 \sqrt{2}} \left(U_0 + \Tilde{U}_1  - \Tilde{U}_2 +
		 \Tilde{U}_3 +V^1 - V^2 + V^3 - \Tilde{V}^0 \right) \\
   &N_2 &:= \frac{1}{ 4 \sqrt{2}} \left(U_1 - U_2  +U_3  - \Tilde{U}_0 -
		 \Tilde{V}^1 +\Tilde{V}^2 - \Tilde{V}^3 - V^0 \right) \\
   &O_3& := 1/4 \left(K + \Tilde{K} + \sqrt{2/3} (R_1 + \Tilde{R}^1 - R_2 -
		 \Tilde{R}^2 - R_3 - \Tilde{R}^3)\right) \\
   &O_1 &:= \frac{1}{ 4 \sqrt{2}} \left(U_0 - \Tilde{U}_1  + \Tilde{U}_2 +
		 \Tilde{U}_3 -V^1 + V^2 + V^3 - \Tilde{V}^0 \right) \\
   &O_2 &:= \frac{1}{ 4 \sqrt{2}} \left(-U_1 + U_2  +U_3  - \Tilde{U}_0 +
		 \Tilde{V}^1 -\Tilde{V}^2 - \Tilde{V}^3 - V^0 \right) \\
   &L_3& := 1/4 \left(K + \Tilde{K} + \sqrt{2/3} (R_1 + \Tilde{R}^1 + R_2 +
		 \Tilde{R}^2 + R_3 + \Tilde{R}^3)\right) \\
   &L_1 &:= \frac{1}{ 4 \sqrt{2}} \left(-U_0 + \Tilde{U}_1  + \Tilde{U}_2 +
		 \Tilde{U}_3 +V^1 + V^2 + V^3 + \Tilde{V}^0 \right) \\
   &L_2 &:= \frac{1}{ 4 \sqrt{2}} \left(U_1 + U_2  +U_3  + \Tilde{U}_0 -
		 \Tilde{V}^1 -\Tilde{V}^2 - \Tilde{V}^3 + V^0 \right)
\end{eqnarray}
They satisfy the commutation relations
\begin{eqnarray}
& \left[M_i, M_j \right] = \epsilon_{ijk} M_k \\
& \left[N_i,N_j \right]  = \epsilon_{ijk}N_k \\
& \left[O_i,O_j \right]  = \epsilon_{ijk}O_k \\
& \left[L_i, L_j \right] = \epsilon_{ijk}L_k
\end{eqnarray}
where the indices $i,j,k$ run from 1 to 3.  The noncompact generators
decompose as 8 doublets under each $SU(2)$ subgroup and all together
form the $(j_L=1/2, j_M=1/2,j_N=1/2, j_O=1/2)$ representation of $
SU(2)_L\times SU(2)_M \times SU(2)_M \times SU(2)_O $.  We shall work
in basis in which the noncompact generators are labelled by the
eigenvalues of $iL_3, iM_3, iN_3$ and $iO_3$ and use the 5-grading
with respect to the eigenvalues of $iL_3$:
\begin{eqnarray}
 \left[ i L_3 , K^{\pm} (m,n,o) \right] &  = &  \pm \frac{1}{2} K^{\pm}(m,n,o)  \\
\left[ i M_3 , K^{\pm} (m,n,o)  \right]  & = & m  K^{\pm}(m,n,o)   \\
 \left[ i N_3 , K^{\pm} (m,n,o) \right] & = & n K^{\pm}(m,n,o)  \\
 \left[ i O_3, K^{\pm} (m,n,o) \right]  &= &  o K^{\pm}(m,n,o)
\end{eqnarray}
where $m, n $ and $o$ take on values $\pm 1/2$.  They are given by the
following linear combinations of the generators defined above
\begin{eqnarray}
 K^-(-1/2,-1/2,-1/2) & = &  ( -iK +i \Tilde{K} + \Delta )  \\
 K^+(1/2,1/2,1/2) & = & ( i K - i \Tilde{K} + \Delta )  \\
 K^-(-1/2,-1/2,1/2) & = & - \frac{1}{2\sqrt{2}} ( -iU_0 -U_1+U_2+U_3+\Tilde{U}_0-i \Tilde{U}_1+i \Tilde{U}_2+\Tilde{U}_3 \notag \\ &&
 -V^0 +iV^1-iV^2-iV^3-i\Tilde{V}^0 -\Tilde{V}^1+\Tilde{V}^2 + \Tilde{V}^3  )  \\
K^+(1/2,1/2,-1/2) & = &- \frac{1}{2\sqrt{2}} (iU_0 -U_1+U_2+U_3+\Tilde{U}_0+i\Tilde{U}_1- i\Tilde{U}_2+\Tilde{U}_3 \notag \\ &&
 -V^0 - iV^1+iV^2+iV^3+i\Tilde{V}^0 -\Tilde{V}^1+\Tilde{V}^2 + \Tilde{V}^3 )  \\
 K^-(-1/2,1/2,1/2) & = &\frac{1}{3} ( i \sqrt{6} R_3 -i \sqrt{6} \Tilde{R}^3 +2 R_3^3 ) \\
 K^+(1/2,-1/2,-1/2) & =& \frac{1}{3} ( - i \sqrt{6} R_3 +i \sqrt{6} \Tilde{R}^3 +2 R_3^3 ) 
 \end{eqnarray}
 \begin{eqnarray} 
 K^-(1/2,1/2,1/2) & =& \frac{1}{2\sqrt{2}} ( -iU_0 +U_1+U_2+U_3 -\Tilde{U}_0 -i \Tilde{U}_1 -i \Tilde{U}_2 -i \Tilde{U}_3 \nonumber \\ &&
 + V^0+iV^1+iV^2+iV^3-i\Tilde{V}^0+\Tilde{V}^1+\Tilde{V}^2+\Tilde{V}^3 ) \\
  K^+(-1/2,-1/2,-1/2) & =& \frac{1}{2\sqrt{2}} ( iU_0 +U_1+U_2+U_3 -\Tilde{U}_0 +i \Tilde{U}_1 +i \Tilde{U}_2 +i \Tilde{U}_3 \nonumber \\ &&
  +V^0-iV^1-iV^2-iV^3+i\Tilde{V}^0+\Tilde{V}^1+\Tilde{V}^2+\Tilde{V}^3 ) \\
  K^-(-1/2,1/2,-1/2) & = &-\frac{1}{2\sqrt{2}} ( -iU_0 +U_1-U_2+U_3 +\Tilde{U}_0 +i \Tilde{U}_1 -i \Tilde{U}_2 +i \Tilde{U}_3 \nonumber \\ &&
 - V^0- iV^1+iV^2-iV^3-i\Tilde{V}^0+\Tilde{V}^1-\Tilde{V}^2 +\Tilde{V}^3 ) \\
  K^+(1/2,-1/2,1/2) & =& -\frac{1}{2\sqrt{2}} ( iU_0 +U_1-U_2+U_3 +\Tilde{U}_0 -i \Tilde{U}_1 +i \Tilde{U}_2 -i \Tilde{U}_3 \nonumber \\ &&
 - V^0+ iV^1-iV^2+iV^3+i\Tilde{V}^0+\Tilde{V}^1-\Tilde{V}^2 +\Tilde{V}^3 ) \\
 K^-(1/2,-1/2,-1/2) & = &-\frac{1}{2\sqrt{2}} ( iU_0 -U_1-U_2+U_3 -\Tilde{U}_0 -i \Tilde{U}_1 -i \Tilde{U}_2 +i \Tilde{U}_3 \nonumber \\ &&
 + V^0+ iV^1+iV^2-iV^3+i\Tilde{V}^0-\Tilde{V}^1-\Tilde{V}^2 +\Tilde{V}^3 ) \\
K^+(-1/2,1/2,1/2) & =& -\frac{1}{2\sqrt{2}} ( -iU_0 -U_1-U_2+U_3 -\Tilde{U}_0 +i \Tilde{U}_1 +i \Tilde{U}_2 -i \Tilde{U}_3 \nonumber \\ &&
 + V^0- iV^1-iV^2+iV^3-i\Tilde{V}^0-\Tilde{V}^1-\Tilde{V}^2 +\Tilde{V}^3 ) \\
 K^-(1/2,-1/2,1/2) & =& - \frac{1}{3} ( i \sqrt{6} R_2 -i \sqrt{6} \Tilde{R}^2  +2R_2^2 ) \\
 K^+(-1/2,1/2,-1/2) & =& - \frac{1}{3} (- i \sqrt{6} R_2 +i \sqrt{6} \Tilde{R}^2  +2R_2^2 ) \\
 K^-(1/2,1/2,-1/2) & =& - \frac{1}{3} ( i \sqrt{6} R_1 -i \sqrt{6} \Tilde{R}^1 +2 R_1^1 ) \\
 K^+(-1/2,-1/2,1/2) & =& - \frac{1}{3} (- i \sqrt{6} R_1 +i \sqrt{6} \Tilde{R}^1 +2 R_1^1 )
  \end{eqnarray}
They satisfy the commutation relations:
\begin{align}
\left[ K^-(m_1,n_1,o_1), K^-(m_2,n_2,o_2) \right] & =  4i (m_1-m_2)(n_1-n_2)(o_1-o_2) L^{(-1)} \\
\left[ K^+(m_1,n_1,o_1), K^+(m_2,n_2,o_2) \right] & =  -4i (m_1-m_2)(n_1-n_2)(o_1-o_2) L^{(+1)} \\
\left[ K^+(m_1,n_1,o_1), K^-(m_2,n_2,o_2) \right] & = - 4i ( (m_1+m_2)(n_1-n_2)(o_1-o_2) M^{(m_1+m_2)}  \notag \\ & + (m_1-m_2)(n_1+n_2)(o_1-o_2) N^{(n_1+n_2)}  \nonumber \\
& + (m_1-m_2)(n_1-n_2)(o_1+o_2) O^{(o_1+o_2)} \notag \\  &- ( m_1-m_2)(n_1-n_2)(o_1-o_2) L_3 \nonumber \\
& - (m_1-m_2)(n_1-n_2) |(o_1-o_2)| O_3  \notag \\ & -  (m_1-m_2)(o_1-o_2) |(n_1-n_2)| N_3 \notag \\ & - (n_1-n_2) (o_1-o_2) |(m_1-m_2)| M_3 )
\end{align}
where
\begin{equation*}
  M^{(\pm1)} = M_1 \pm i M_2
\end{equation*}
\begin{equation*}
  N^{(\pm1)} = N_1 \pm i N_2
\end{equation*}
\begin{equation*}
  O^{(\pm 1)} =O_1 \pm i O_2
\end{equation*}

\subsection{ Spherical Vector of Quasiconformal Action of $SO(4,4)$}
\setcounter{equation}{0}

Unitary representations induced by the quasiconformal action with
unitary character $\nu$ include the quaternionic discrete series
representations of Gross and Wallach \cite{MR1421947} as was shown
explicitly for rank two cases in \cite{Gunaydin:2007qq}. Critical to
the analysis of \cite{Gunaydin:2007qq} is the explicit expression for
the spherical vector of quasiconformal realizations of $SU(2,1)$ and
$G_{2(2)}$. The quaternionic discrete series representations and their
continuations appear as submodules in the Verma modules generated by
the action of  noncompact generators on the spherical vectors for special values of the
parameter $\nu$.
To carry out this program for $SO(4,4)$ and higher groups we need to
determine the spherical vector of their quasiconformal
realizations. The spherical vectors can, in principle, be obtained
by solving the differential equations
\begin{equation}
C_M \Phi_{\nu}(p,q,x)=0
\end{equation}
where $C_M$ denote the compact generators in the quasiconformal
realization. This is however quite unwieldy considering the fact that
the relevant differential operators are highly nonlinear and involve
9 variables for $SO(4,4)$. The situation gets much worse for the
quasiconformal groups of higher Jordan algebras. However, we were
able to deduce the spherical vector of $SO(4,4)$ from that of
$G_{2(2)}$ given in \cite{Gunaydin:2007qq} simply by using properties
of Jordan algebras of degree three and the fact that the
quasiconformal realization of $G_{2(2)}$ can be obtained from that of
any Jordan algebra of degree three by restricting the Jordan algebra
to its identity element.

We find that the spherical vector of $SO(4,4)$ is given simply by
\begin{eqnarray}
\Phi_{\nu}(p,q,x) &=& [ (1+x^2)^2 +( I_4)^2  +8 I_4 -2(1+x^2)(I_4-I_2)  \nn \\ && + \frac{1}{2} J_6 + 8 x J_4+ \frac{4}{81} H_4 ]^{\frac{\nu}{4}}
\end{eqnarray}
where
\begin{eqnarray}
I_4 & = 4p^0q_1q_2q_3 + 4q_0 p^1p^2p^3 + (p^0q_0-p^1q_1-p^2q_2-p^3q_3)^2 \\ \nonumber & -4(p^2q_2p^3q_3 +p^1q_1p^2q_2+p^1q_1p^3q_3)  \nonumber
\end{eqnarray}
\begin{equation}
I_2 = (p^0)^2 + (q_0)^2 + p^I p^I + q_I q_I
\end{equation}
\begin{equation}
J_4= \frac{1}{4} \left(p^0 \frac{\partial{I_4}}{\partial{q_0}} - q_0 \frac{\partial{I_4}}{\partial{p^0}}
+q_I \frac{\partial{I_4}}{\partial{p^I}} - p^I \frac{\partial{I_4}}{\partial{q_I}} \right)
\end{equation}
\begin{equation}
J_6 = \left( \frac{\partial{I_4}}{\partial{p^0}}\right)^2 +\left( \frac{\partial{I_4}}{\partial{q_0}}\right)^2 +\left( \frac{\partial{I_4}}{\partial{p^I}}\right)
\left( \frac{\partial{I_4}}{\partial{p^I}}\right) + \left( \frac{\partial{I_4}}{\partial{q_I}}\right)
\left( \frac{\partial{I_4}}{\partial{q_I}}\right)
\end{equation}
\begin{equation}
H_4 = 27 \left( (p^{\#})_I -\sqrt{3} q_0 p^I - \sqrt{3} p^0 q_I + (q_{\#})^I \right)\left( (p^{\#})_I -\sqrt{3} q_0 p^I - \sqrt{3} p^0 q_I + (q_{\#})^I \right)
\end{equation}
where $ (p^{\#})_I= C_{IJK}p^Jp^K $ and $ (q_{\#})^I= C^{IJK}q_Jq_K $
and $I,J,K,...=1,2,3$

The generators $C_M$ of the maximal compact subgroup $SO(4)\times
SO(4)$ all annihilate the vector $\Phi_{\nu}(p,q,x)$ for arbitrary
values of $\nu$. Unitary irreducible representations, including the
quaternionic discrete series induced by the above quasiconformal
realization starting from the spherical vector will be studied
elsewhere \cite{mgop}.

\section{Spherical Vectors  of Quasiconformal Groups associated with general Euclidean  Jordan algebras of Degree Three}
\renewcommand{\theequation}{\arabic{section}.\arabic{equation}}
\setcounter{equation}{0}

Every Jordan algebra $J$ of degree three admits three mutually orthogonal irreducible idempotents $\mathbb{P}_1,\mathbb{P}_2,\mathbb{P}_3$:
\begin{equation}
  \mathbb{P}_i \circ \mathbb{P}_j = \delta_{ij} \mathbb{P}_i
\end{equation}
\begin{equation}
  Tr(\mathbb{P}_i) = 1 \hspace{1cm} i,j,..=1,2,3 \nonumber
\end{equation}
By the action $U$ of the automorphism group $Aut(J)$ one can bring a
general element $X$ of the Jordan algebra to the form\footnote{As
stated above we shall work in a basis $e_I$ where the three
irreducible idempotents are the first three basis elements $\mathbb{P}_i = e_i$
for $i=1,2,3$. The parameters $\lambda_i$ are the analogs of the
light-cone coordinates in Minkowskian spacetimes.}
\begin{equation}
  Aut(J): \;\;\; X \longrightarrow U X U^{-1}= \lambda_1 \mathbb{P}_1 + \lambda_2 \mathbb{P}_2 + \lambda_3 \mathbb{P}_3
\end{equation}
such that its norm is simply
\begin{equation}
  \mathcal{N}(X) =\lambda_1 \lambda_2 \lambda_3
\end{equation}

For simple Jordan algebras $J_3^{\mathbb{R}}$, $J_3^{\mathbb{C}}$ and
$J_3^{\mathbb{H}}$ this corresponds simply to diagonalizing a $3\times
3$ symmetric, complex Hermitian and quaternionic Hermitian matrix by
an $SO(3)$, $SU(3)$ and $USp(6)$ transformation. For the exceptional
Jordan algebra $J_3^{\mathbb{O}}$ the diagonalization procedure is
more subtle and can be achieved by an $F_4$ transformation
\cite{Gunaydin:1978jq}. For the generic Jordan family $J= \mathbb{R}
\oplus \Gamma_{(1,n-1)}$, by the action of the automorphism group
$SO(n-1)$ one can rotate the cubic norm of a general element to that
of an element of subalgebra $ \mathbb{R} \oplus \Gamma_{(1,1)}$ whose
norm can be written in the above form as we saw in the previous
section. Thus the general quasiconformal group actions on a Jordan
algebra of degree three can be transformed into the action of
$SO(4,4)$ of previous section.

To construct the spherical vector of a general quasiconformal group
starting from that of $SO(4,4)$ we will use the fact that closure of
the Lie algebra $SO(4,4)$ and Lie algebra of automophism group
$Aut(J)$ of the Euclidean Jordan algebra $J$ is the full
quasiconformal Lie algebra $QCon(\mathcal{F}(J))$. Similarly the
closure of Lie algebra of maximal compact subgroup $SO(4)\times SO(4)$
of $SO(4,4)$ and Lie algebra of $Aut(J)$ is the Lie algebra of maximal
compact subgroup $\widetilde{Conf}(J) \times SU(2)$ of the
quasiconformal group $QConf(J)$. Written in terms of the
C-tensor (or equivalently the cubic norm) the extension of the terms
in the spherical vector of $SO(4,4)$ to those that occur in the
spherical vector of general $QConf(J)$ is straightforward. However, one
complication is the fact that there are terms in the spherical vector
of general $QCon(J)$ that vanish when restricted to $SO(4,4)$
subalgebra. One writes down an Ansatz for such terms with arbitrary coefficients and then determines these
coefficients by the requirement of invariance under the maximal
compact subgroup $\widetilde{Conf}(J) \times SU(2)$.

We find that the spherical vector of a general quasiconformal group  $QConf(J)$ associated with a Euclidean Jordan algebra $J$ is given by
\begin{eqnarray}
    \Phi_{\nu}(p,q,x) &=& [ (1+x^2+ I_2 -I_4)^2 - (I_2)^2  +8 I_4
      + \frac{1}{2} I_6 + 8 x J_4 + \frac{4}{81} H_4 ]^{\frac{\nu}{4}}
\end{eqnarray}
where
\begin{equation}
   I_2 = (p^0)^2 + (q_0)^2 + p^I p^I + q_I q_I
\end{equation}
\begin{eqnarray}
I_4 &=&
    \left(p^0 q_0 -  p^I q_I \right)^2 - \frac{4}{3}  C_{IJK} p^J p^K C^{ILM} q_L q_M \\  \nonumber && +
             \frac{4}{3\sqrt{3}} p^0 C^{IJK} q_I q_J q_K + \frac{4}{3\sqrt{3}} q_0 C_{IJK} p^I p^J p^K
\end{eqnarray}
\begin{equation}
      J_4 = \frac{1}{4} \left(p^0 \frac{\partial{I_4}}{\partial{q_0}} -
                              q_0 \frac{\partial{I_4}}{\partial{p^0}}
			      +q_I \frac{\partial{I_4}}{\partial{p^I}} -
			      p^I \frac{\partial{I_4}}{\partial{q_I}}
                        \right)
\end{equation}
\begin{equation}
   I_6 = \left( \frac{\partial{I_4}}{\partial{p^0}}\right)^2 +\left(
       \frac{\partial{I_4}}{\partial{q_0}}\right)^2 +\left(
       \frac{\partial{I_4}}{\partial{p^I}}\right) \left(
       \frac{\partial{I_4}}{\partial{p^I}}\right) + \left(
       \frac{\partial{I_4}}{\partial{q_I}}\right) \left(
       \frac{\partial{I_4}}{\partial{q_I}}\right) + 4I_4 I_2
\end{equation}
\begin{eqnarray}
    H_4 & = &27 \left( (p^{\#})_I -\sqrt{3} q_0 p^I - \sqrt{3} p^0 q_I
    + (q_{\#})^I \right)\left( (p^{\#})_I -\sqrt{3} q_0 p^I - \sqrt{3}
    p^0 q_I + (q_{\#})^I \right) \nonumber \\ && + C_4(J)
\end{eqnarray}
where $ (p^{\#})_I= C_{IJK}p^Jp^K $ and $ (q_{\#})^I= C^{IJK}q_Jq_K $.
 $C_4(J)$ is the ``correction'' term that vanishes when restricted
to the subalgebra $SO(4,4)$ and has a different form for simple
Jordan algebras and non-simple ones.  For simple Euclidean Jordan
algebras of degree three the quartic correction term $C_4(J)$ is given by
\begin{eqnarray}
      C_4 (J_3^{\mathbb{A}}) & = & 81 \left( \mathop{Tr}\left[ M_0(p)\circ M_0(q)\right]\right)^2 +
                \frac{81}{2} \mathop{Tr} \left[ M_0(p)^2 \right] \mathop{Tr} \left[ M_0(q)^2 \right] \\ \nonumber &&
		-243 \mathop{Tr} \left[\{M_0(p), M_0(q), M_0(p)\}\circ M_0(q) \right]
\end{eqnarray}
where
\begin{eqnarray}
 M_0(q)  &=& M(q) - \frac{1}{3} Tr M(q) \\   M(q) & = & e^I q_I \in J_3^{\mathbb{A}}
 \end{eqnarray}
and similarly for $M_0(p)$.
$\{A,B,C\}$ denotes the Jordan triple product
\begin{equation}
\{A,B,C\} = A\circ ( B \circ C ) + C \circ ( B \circ A ) - (A\circ C) \circ B
\end{equation}
For special Jordan algebras with the Jordan product
\begin{equation*}
   A \circ B = \frac{1}{2} (A B + B A )
\end{equation*}
we have
\begin{equation}
   \left\{ A,B,A \right\} = A B A
\end{equation}
Hence for Jordan algebras $J_3^{\mathbb{R}}, J_3^{\mathbb{C}}$ and $J_3^{\mathbb{H}}$ the term $C_4(J)$ can be written as
\begin{eqnarray}
          C_4 (J_3^{\mathbb{A}}) &= &
             81 \left(\mathop{Tr} \left[ M_0(p) M_0(q) \right] \right)^2 +
             \frac{81}{2} \mathop{Tr} \left[ M_0(p)^2 \right] \mathop{Tr} \left[ M_0(q)^2 \right] \\ \nonumber
             && -243 \mathop{Tr} \left[M_0(p) M_0(q) M_0(p) M_0(q) \right] \nonumber
\end{eqnarray}
We recall that the basis elements $e_I$ of $J_3^{\mathbb{A}}$ are normalized such that
\begin{eqnarray*}
  \mathop{Tr} (e_I \circ e_J ) = \delta_{IJ} \\
\mathop{Tr} (\Tilde{e}^I \circ \Tilde{e}^J ) = \delta^{IJ}
\end{eqnarray*}
and
\begin{equation*}
\mathcal{N}(M) = 3 \sqrt{3} Det M
\end{equation*}
For $J_3^{\mathbb{R}} $ we label our basis elements such that
\begin{equation}
M (p)  = \frac{1}{\sqrt{2}}\left( {\begin{array}{*{20}c}
   \sqrt{2} p^1 & p^6 & p^5  \\
   p^6 & \sqrt{2}p^2 & p^4  \\
   p^5 & p^4 &\sqrt{2}p^3  \\
\end{array}} \right)
\end{equation}
and
\begin{equation}
       \mathcal{N}(M(p)) = 3 \sqrt{3} \left\{ p^1 p^2 p^3 -
           \frac{1}{2} \left[  p^1 (p^4)^2 + p^2 (p^5)^2 + p^3 (p^6)^2 \right]
         + \frac{1}{\sqrt{2}} p^4 p^5 p^6 \right\}
\end{equation}

For the Jordan algebras $J_3^{\mathbb{A}}$, where $\mathbb{A}=
\mathbb{C}$, $\mathbb{H}$ or $\mathbb{O}$ coordinates $p^4, p^5$ and $p^6 $ become
elements of $\mathbb{A}$, which we will denote by capital letters
$P^4, P^5$ and $P^6$. Thus for $M(p) \in J_3 ^{\mathbb{A}}$ we have
\begin{equation}
M (p)  = \frac{1}{\sqrt{2}}\left( {\begin{array}{*{20}c}
   \sqrt{2} p^1 & P^6 & \bar{P}^5  \\
   \bar{P}^6 & \sqrt{2}p^2 & P^4  \\
   P^5 & \bar{P}^4 &\sqrt{2}p^3  \\
\end{array}} \right)
\end{equation}
and the cubic norm of $M(p)$ becomes
\begin{equation}
    \mathcal{N}(M(p)) = 3 \sqrt{3}\{ p^1 p^2 p^3 -
        \frac{1}{2} \left( p^1 |P^4|^2 + p^2 |P^5|^2 + p^3 |P^6|^2 \right)
        + \frac{1}{\sqrt{2}} Re (P^4 P^5 P^6) \}
\end{equation}
where $Re(X)$ denotes the real part of $X \in \mathbb{A}$ and $|X|^2 = X \bar{X} $.
If we expand the elements $P^4, P^5 $ and $P^6$ in terms of their real components
\begin{eqnarray}
 P^4 = p^4 + p^{4+3A} j_A  \nonumber \\
 \bar{P}^4 = p^4 - p^{4+3A} j_A \nonumber \\
 P^5 = p^5 + p^{5+3A} j_A \nonumber \\
 \bar{P}^5 = p^5 - p^{5+3A} j_A \\
 P^6 = p^6 + p^{6+3A} j_A \nonumber \\
 \bar{P}^6 = p^6 - p^{6+3A} j_A \nonumber
\end{eqnarray}
where the index $A$ is summed over and using the fact that the imaginary units satisfy
\begin{equation}
 j_A j_B = - \delta_{AB} + \eta_{ABC} j_C
\end{equation}
we can write the cubic norm as
\begin{eqnarray}
      \mathcal{N}(M(p))& = &3 \sqrt{3} \{ p^1 p^2 p^3 -  \frac{1}{2} p^1 [( p^4)^2  + p^{4+3A} p^{4+3A}] \\ &&
       -\frac{1}{2} p^2 [ (p^5)^2 + p^{5+3A} p^{5+3A} ] -\frac{1}{2} p^3 [ (p^6)^2 + p^{6+3A} p^{6+3A} ] \nonumber  \\ &&
      + \frac{1}{\sqrt{2}} [ p^4 p^5 p^6   - p^4 p^{(5+3A)} p^{(6+3A)} - p^5 p^{(4+3A)} p^{(6+3A)} -
	p^6 p^{(4+3A)}  p^{(5+3A)} ] \nonumber \\ &&
      -\frac{1}{\sqrt{2}} \eta_{ABC} p^{4+3A} p^{5+3B} p^{6+3C} \nonumber \}
\end{eqnarray}
The indices $A,B,C$ take on the single value $1$ for $\mathbb{C}$, run
from $1$ to $3$ for $\mathbb{H}$ and from $1$ to $7$ for
$\mathbb{O}$. Note that $\eta_{ABC}$ vanishes for $\mathbb{C}$.

For the generic nonsimple Jordan algebras $(\mathbb{R} \oplus \Gamma_{(1,n-1)})$  of degree three the cubic form is
\begin{equation}
\mathcal{N}(q) =C^{IJK} q_Iq_Jq_K=  \frac{3\sqrt{3}}{2} q_1 [ 2q_2q_3 - (q_4)^2 -(q_5)^2- \cdots -(q_{n+1})^2 ]
\end{equation}
and the quartic correction term $C_4$ that appears in  $H_4$ is given by
\begin{eqnarray}
        C_4 ( \mathbb{R}\oplus \Gamma_{(1,n-1)})  &= &
      - \frac{81}{2} \left\{ (p^2 - p^3)(q_2 - q_3) + 2 p^4 q_4 + \cdots +2 p^{n+1} q_{n+1} \right\}^2  \\ &&
    + \frac{81}{2} \left\{ (p^2 -p^3)^2 + 2 (p^4)^2 + \cdots + 2 (p^{n+1})^2 \right\}
     \left\{ (q_2 - q_3)^2 + 2 (q_4)^2 + \cdots + 2 (q_{n+1})^2 \right\} \nonumber
\end{eqnarray}

\section{ Exceptional Groups $E_{6(6)}$, $E_{7(7)}$, $E_{8(8)}$ and  $SO(m+4,n+4)$ as Quasiconformal Groups  and Non-Euclidean Jordan Algebras  }
\setcounter{equation}{0}

Jordan algebras of degree three that define $N=2$ MESGTs in $d=5$ with
symmetric target spaces are all Euclidean and the quasiconformal
groups associated with them are of the quaternionic real form that we
studied in the previous section. The general formulas given for the
quasiconformal Lie algebra in a basis covariant with respect to the
reduced structure group of the Jordan algebra are valid for all real
forms of the underlying Jordan algebras of degree three.
Split exceptional group $E_{8(8)}$ is the U-duality group of maximal
supergravity in three dimensions and its quasiconformal realization
was first given in \cite{Gunaydin:2000xr}, in a basis covariant with
respect to the four dimensional U-duality group $E_{7(7)}$. By
truncation the quasiconformal realizations of $E_{7(7)}$, $E_{6(6)}$
and $F_{4(4)}$ were obtained in a basis covariant with respect to
$SO(6,6), SL(6,\mathbb{R}) $ and $Sp(6,\mathbb{R})$, respectively
\cite{Gunaydin:2000xr}.  The groups $E_{7(7)}, SO(6,6),
SL(6,\mathbb{R})$ and $Sp(6,\mathbb{R})$ are the conformal groups of
split Jordan algebras of $J_3^{\mathbb{O}_s}, J_3^{\mathbb{H}_s},
J_3^{\mathbb{C}_s}$ and $J_3^{\mathbb{R}}$, respectively.

To obtain the quasiconformal realizations of split exceptional groups
in a basis covariant with respect to Lorentz (reduced structure)
groups of underlying Jordan algebras we simply need to substitute in
the formulas of section 4 cubic norm forms (or C-tensors) of simple
split Jordan algebras in place of those of Euclidean Jordan algebras.

The quasiconformal group associated with $J_3^{\mathbb{R}}$ is the
split exceptional group $F_{4(4)}$ whose quotient with respect to its
maximal compact subgroup $USp(6)\times USp(2)$ is a quaternionic
symmetric space. In other words for $J_3^{\mathbb{R}}$ the
quasiconformal group is both split and quaternionic real.
However if we replace the underlying division algebras of the other
simple Jordan algebras of degree three to be of the split form
$\mathbb{C}_s$, $\mathbb{H}_s$ and $\mathbb{O}_s$, then the resultant
quasiconformal groups are no longer quaternionic real. They yield the
quasiconformal realizations of split exceptional groups $E_{6(6)}$,
$E_{7(7)}$ and $E_{8(8)}$, respectively, in a basis covariant with
respect to their reduced structure groups $SL(3,\mathbb{R})\times
SL(3,\mathbb{R})$, $SL(6,\mathbb{R})$ and $E_{6(6)}$, respectively.

Similarly, in the formulas of section four if we replace the norm
forms (C-tensors) of non-simple Euclidean Jordan algebras $(
\mathbb{R} \oplus \Gamma_{(1,m)})$ with those of non-Euclidean Jordan
algebras $ (\mathbb{R} \oplus \Gamma_{(n,m)})$ $(n \neq 1)$ we get
the quasiconformal realizations of groups $SO(n+3,m+3)$ covariant with
respect their reduced structure groups $SO(1,1) \times SO(n,m)$.

The formulas for spherical vectors of the quasiconformal groups
associated with euclidean Jordan algebras do not however carry over
directly to the spherical vectors of quasiconformal groups of
non-Euclidean Jordan algebras and will be studied elsewhere.

{\bf Acknowledgement:}
We would like thank Andy Neitzke, Boris Pioline and Andrew Waldron
for discussions in the early stages of this work. One of us (M.G)
would like to acknowledge the hospitality extended to him during his
sabbatical stay at the Institute for Advanced Study, Princeton,
where part of this work was done and support of Monell Foundation
during his stay.  This work was supported in part by the National
Science Foundation under grant number PHY-0555605. Any opinions,
findings and conclusions or recommendations expressed in this
material are those of the authors and do not necessarily reflect the
views of the National Science Foundation.

\section{Appendices}

\appendix
\setcounter{equation}0
\def\theequation{A.\arabic{equation}}
\section{ Euclidean and Split Jordan algebras of Degree Three}
Referring to the recent monograph  \cite{MR2014924} for references and details we shall give a brief review of Jordan algebras in this Appendix.
A Jordan algebra over a field $\mathbb{F}$ is an algebra, $J$ with a
symmetric product $\circ$
\begin{equation}\label{commute} X\circ Y = Y
\circ X \in J, \quad \forall\,\, X,Y \in J \ ,
\end{equation}
that satisfies the Jordan identity
\begin{equation}\label{Jidentity}
X\circ (Y \circ X^2)= (X\circ Y) \circ X^2 \ ,
\end{equation}
where $X^2\equiv (X\circ X)$.  Hence a Jordan algebra is commutative
and in general not associative algebra.

Given a Jordan algebra $J$, one can define a norm form,
$\mathbf{N}:J\rightarrow \mathbb{R}$ over it that satisfies the composition
property \cite{MR0251099}
\begin{equation}\label{Norm}
\mathbf{N}[2X\circ(Y\circ X)-(X\circ X)\circ Y]=\mathbf{N}^{2}(X) \mathbf{N}(Y).
\end{equation}
The degree, $p$, of the norm form as well as of $J$ is defined by
$\mathbf{N}(\lambda X)=\lambda^p \mathbf{N}(X)$, where $\lambda\in \mathbb{R}$.  A
\emph{Euclidean} Jordan algebra is a Jordan algebra for which the
condition $X\circ X + Y\circ Y=0$ implies that $X=Y=0$ for all $X,Y\in
J$. They are sometimes called compact Jordan algebras since the
automorphism groups of Euclidean Jordan algebras are compact.

As was shown in \cite{Gunaydin:1983bi}, given a Euclidean Jordan
algebra of degree three one can identify its norm form $\mathbf{N}$ with the
cubic polynomial $\mathcal{V}$ defined by the C-tensor of a 5D, $N=2$
MESGT with a symmetric scalar manifold.
 Euclidean Jordan algebras of degree three fall into an infinite family of non-simple Jordan algebras of the form
\begin{equation*}
  J=\mathbb{R}\oplus \Gamma_{(1,n-1)}
\end{equation*}
 which is referred to as  the generic
Jordan family. The scalar manifolds of corresponding 5D, $N=2$ MESGT's are
\begin{equation*}
 \mathcal{M} =\frac{SO(n -1,1)}{SO(n -1)}\times  SO(1,1)
\end{equation*}
$\Gamma_{(1,n-1)}$ is an $n$
dimensional Jordan algebra of degree two associated with a quadratic
norm form in $n$ dimensions that has a ``Minkowskian signature''
$(+,-,\ldots,-)$.  A simple realization of $\Gamma_{(1,n-1)}$ is provided by
$(n-1)$ Dirac gamma matrices $\gamma^i$ $(i,j,\ldots=1,\ldots,(n-1))$
of an $(n-1)$ dimensional Euclidean space together with the identity
matrix $ \gamma^0 = \mathbf{1}$ and the Jordan product $\circ$ being
one half the anticommutator:

\begin{eqnarray}
\gamma^i \circ \gamma^j &=& \frac{1}{2}
\{\gamma^i,\gamma^j\}= \delta^{ij} \gamma^0
\nonumber \\
\gamma^0 \circ \gamma^0 &=& \frac{1}{2}
\{\gamma^0,\gamma^0\}= \gamma^0 \nonumber\\
\gamma^i \circ \gamma^0 &=& \frac{1}{2}
\{\gamma^i,\gamma^0\}= \gamma^i \ .
\end{eqnarray}
  The quadratic norm of a general element $\mathbb{X} = X_0 \gamma^0 + X_i \gamma^i $ of
$\Gamma_{(1,n-1)}$
is defined as
\begin{equation*}
  \mathbf{Q}(\mathbb{X}) = \frac{1}{2^{[n/2]}} \mathop{Tr}
  \mathbb{X} \bar{\mathbb{X}} = X_0X_0 - X_iX_i \ ,
\end{equation*}
where
\begin{equation*}  \bar{\mathbb{X}} \equiv  X_0 \gamma^0 - X_i \gamma^i  \ .
\end{equation*}
The norm of a general element $y \oplus \mathbb{X} $ of the non-simple
Jordan algebra $J=\mathbb{R}\oplus \Gamma_{(1,n-1)}$ is
simply given by
\begin{equation}
  \mathbf{N}(y \oplus \mathbb{X}) =  y \mathbf{Q}(\mathbb{X})
\end{equation}
where $y\in \mathbb{R}$.

There exist four simple Euclidean Jordan algebras of degree three.
They are generated by Hermitian $(3\times 3)$-matrices over the four
division algebras $\mathbb{A} = \mathbb{R}, \mathbb{C},
\mathbb{H}, \mathbb{O}$
\begin{equation*}
 J =
\left(
  \begin{array}{ccc}
    \alpha & Z & \bar{Y} \\
    \bar{Z} & \beta & X \\
    Y & \bar{X} & \gamma \\
  \end{array}
\right)
\end{equation*}
where $\alpha , \beta, \gamma \in \mathbb{R}$ and $X,Y,Z \in
\mathbb{A}$ with the product being one half the anticommutator.  They
are denoted as $J_3^{\mathbb{R}}$, $J_3^{\mathbb{C}}$,
$J_3^{\mathbb{H}}$, $J_3^{\mathbb{O}}$, respectively, and the
corresponding $N=2$ MESGts are called ``magical supergravity
theories''.  They have the 5D scalar manifolds:
\begin{eqnarray}\label{magicals}
J_{3}^{\mathbb{R}}:\quad \mathcal{M}&=& SL(3,\mathbb{R})/
SO(3)\qquad
\nonumber\\
J_{3}^{\mathbb{C}}:\quad \mathcal{M}&=& SL(3,\mathbb{C})/
SU(3)\qquad
\nonumber\\
J_{3}^{\mathbb{H}}:\quad \mathcal{M}&=& SU^{*}(6)/
USp(6)\qquad
 \nonumber \\
J_{3}^{\mathbb{O}}:\quad \mathcal{M}&=& E_{6(-26)}/
F_{4}\qquad \qquad
\\ .
\end{eqnarray}

The cubic norm form, $\mathbf{N}$, of these Jordan algebras is given by the
determinant of the corresponding Hermitian $(3\times 3)$-matrices
(modulo an overall scaling factor).
\begin{equation}
\mathbf{N}(J)= \alpha \beta \gamma - \alpha X \bar{X} - \beta Y \bar{Y} - \gamma Z \bar{Z} + 2 Re (X Y Z )
\end{equation}
where $Re(XYZ)$ denotes the real part of $XYZ$ and bar denotes conjugation in the underlying division algebra.

For a real quaternion $X \in \mathbb{H}$ we have
\begin{eqnarray}
X&=&X_0 + X_1 j_1 + X_2 j_2 + X_3 j_3 \nonumber \\
\bar{X}& = &X_0 - X_1 j_1 - X_2 j_2 - X_3 j_3 \\
X \bar{X}& =& X_0^2 + X_1^2 + X_2^2 +X_3^2 \nonumber
\end{eqnarray}
where the imaginary units $j_i$ satisfy
\begin{equation}
j_i j_j = - \delta_{ij} + \epsilon_{ijk} j_k
\end{equation}
For a real  octonion $X\in \mathbb{O}$ we have
\begin{eqnarray}
X &=& X_0 + X_1 j_1 + X_2 j_2 + X_3 j_3  + X_4 j_4 + X_5 j_5  +X_6 j_6 + X_7 j_7 \nonumber \\
\bar{X} &=& X_0 - X_1 j_1 - X_2 j_2 - X_3 j_3  - X_4 j_4 - X_5 j_5  - X_6 j_6 - X_7 j_7 \\
X \bar{X} &= &X_0^2 + \sum_{A=1}^7 (X_A)^2 \nonumber
\end{eqnarray}
Seven imaginary units of real  octonions satisfy
\begin{equation}
j_A j_B = - \delta_{AB} + \eta_{ABC} j_C
\end{equation}
where $\eta_{ABC}$ is completely antisymmetric and in the conventions of \cite{Gunaydin:1973rs} take on the values
\begin{equation}
\eta_{ABC} = 1  \Leftrightarrow  (ABC)= (123), (471), (572), (673), (624), (435), (516)
\end{equation}

In the generic infinite family of non-simple Jordan algebras of degree
three , $ \mathbb{R} \oplus \Gamma$, one can take the
quadratic form defining  the Jordan algebra $\Gamma$ of degree two to be of
arbitrary signature different from Minkowskian, which result in
non-compact or non-Euclidean Jordan algebras. If the quadratic norm
form has signature $(p,q)$ we shall denote the Jordan algebra as
$\Gamma_{(n,m)}$. It is generated by $(n+m-1)$ Dirac gamma matrices
$\gamma_\mu$ satisfying
\begin{equation}
  \{ \gamma_\mu , \gamma_\nu \} = 2 \eta_{\mu\nu} \gamma^0
\end{equation}
where $\eta_{\mu\nu}$ has signature $(n-1,m)$. Then the quadratic norm
is invariant under $ SO(n,m)$ and the reduced structure group of the
Jordan algebra $ \mathbb{R} \oplus \Gamma_{(n,m)}$ is
\begin{equation*} SO(n,m) \times SO(1,1) \end{equation*}
The invariant tensor $C_{IJK}$ defining the cubic norm of the Jordan algebra
$(\mathbb{R}\oplus \Gamma_{(n,5)} ) $ can be identified with the
invariant tensor in 5D $N=4$ MESGT's describing the $F \wedge
F \wedge A $ coupling of $n$ , $N=4$ vector multiplets coupled to
$N=4$ supergravity. Scalar manifolds of these theories are symmetric
spaces
\begin{equation}
SO(p,5)\times SO(1,1) /SO(5) \times SO(p)
\end{equation}

Scalar manifold of $N=6$ supergravity is the same as that of $N=2$
MESGT defined by the simple Jordan algebra
$J_3^{\mathbb{H}}$\cite{Gunaydin:1983rk}, namely
\begin{equation*} SU^*(6)/USp(6) \end{equation*}
Therefore its invariant C-tensor is simply the one given by the cubic
norm of $J_3^{\mathbb{H}}$.

As for N=8 supergravity in five dimensions its C-tensor is simply the
one given by the cubic norm of the split exceptional Jordan algebra
$J_3^{\mathbb{O}_s}$ defined over split octonions $\mathbb{O}_s$. Four
of the seven ``imaginary units'' of split octonions square to $+1$. If
we denote the split imaginary units as $j_{\mu}^s$ ($\mu =4,5,6,7$)
and the imaginary units of the real quaternion subalgebra as $j_i , (
i=1,2,3)$ we have:
\begin{eqnarray}
j_{\mu}^s j_{\nu}^s& =& \delta_{\mu \nu} - \eta_{\mu \nu i} j_i \nonumber \\
j_i j_j &=& -\delta_{ij} + \epsilon_{ijk} j_k \\
j_i j^s_{\mu} &=& \eta_{i\mu\nu} j^s_\nu \nonumber
\end{eqnarray}
where $\eta_{ABC}$  $(A,B,C=1,2..7)$ are the structure constant of the real octonion algebra $\mathbb{O}$ defined above.
For a split octonion
\begin{equation*}
  O_s = o_0 + o_1 j_1 + o_2 j_2 + o_3 j_3  + o_4 j^s_4 + o_5 j^s_5  +o_6 j^s_6 + o_7 j_7^s
\end{equation*}
the norm is
\begin{equation*} O_s \bar{O}_s = o_0^2 +o_1^2 + o_2^2 +o_3^2 - o_4^2 -o_5^2 -o_6^2 -o_7^2 \end{equation*}
where $\bar{O}_s = o_0 + o_1 j_1 + o_2 j_2 + o_3 j_3 - o_4 j^s_4 - o_5
j^s_5 - o_6 j^s_6 - o_7 j_7^s $. The norm has the invariance group
$SO(4,4)$. The automorphism group of the split exceptional Jordan
algebra defined by $ 3\times 3 $ Hermitian matrices of the form
\begin{equation}
J^s =\left(
       \begin{array}{ccc}
         \alpha & Z^s& \bar{Y}^s\\
         \bar{Z}^s & \beta & X^s \\
         Y^s & \bar{X}^s & \gamma \\
       \end{array}
     \right)
\end{equation}
is the noncompact group $F_{4(4)}$ and its reduced structure group is $E_{6(6)}$ under which the C-tensor is invariant. $E_{6(6)}$ is  the invariance group of maximal supergravity  in five dimensions whose  scalar manifold is
\begin{equation*} E_{6(6)} /USp(8) \end{equation*}

The split quaternion algebra $\mathbb{H}^s$ has two ``imaginary units'' $j^s_m$ (m=2,3) that square to +1:
\begin{eqnarray}
j^s_m j^s_n = \delta_{mn} - \epsilon_{mnk} j_k \\
(j_1)^2 =-1 \nonumber \\
j_1 j^s_m = \epsilon_{1mn} j_n^s \nonumber
\end{eqnarray}
For a split quaternion
\begin{equation*}
  Q_s = q_0 + q_1 j_1 + q_2 j_2^s + q_3 j^s_3
\end{equation*}
the norm is
\begin{equation*}
     Q_s \bar{Q}_s = q_0^2 +q_1^2 -q_2^2 -q_3^2
\end{equation*}
where $\bar{Q}_s = q_0 - q_1 j_1 -q_2 j_2^s -q_3 j^s_3 $ and it is
invariant under $SO(2,2)$.  The automorphism group of the split Jordan
algebra $J_3^{\mathbb{H}_S}$ is $Sp(6,\mathbb{R})$ with the maximal
compact subgroup $SU(3)\times U(1)$. Its reduced structure group is
$SL(6,\mathbb{R})$.

The split complex numbers have an ``imaginary unit'' that squares to +1
and its norm has $SO(1,1)$ invariance. The automorphism group of split
complex Jordan algebra $J_3^{\mathbb{C}_s}$ is $SL(3,\mathbb{R})$ and
its reduced structure group is
\begin{equation*}
  SL(3,\mathbb{R})\times SL(3,\mathbb{R})
\end{equation*}

\section{Conformal Groups of Jordan Algebras}
\def\theequation{B.\arabic{equation}}
\setcounter{equation}{0}
 Generalized conformal group $Conf(J)$  of a Jordan algebra $J$ is
generated by translations $T_{\mathbf{a}}$ , special conformal generators $K_{\mathbf{a}}$ ,  Lorentz transformations and dilatation generators $M_{\mathbf{a}\mathbf{b}}$.
Lorentz transformations and dilatations generate  the structure algebra $\mathfrak{str}(J)$  of $J$.
\cite{Gunaydin:1975mp,Gunaydin:1989dq,Gunaydin:1992zh}.  Lie algebra $\mathfrak{conf}(J)$ of the conformal group $Conf(J)$ has  a 3-grading
with respect to the generator $D$ of dilatations:
\begin{equation}
  \mathfrak{conf}(J) = K_{\mathbf{a}} \oplus M_{\mathbf{a}\mathbf{b}} \oplus T_{\mathbf{b}}
\end{equation}

The conformal Lie algebra $\mathfrak{conf}(J)$ acts on the elements $\mathbf{x}$ of a Jordan algebra $J$    as follows:
\begin{eqnarray}
T_{\mathbf{a}} \mathbf{x}  = \mathbf{a} \nonumber \\
M_{\mathbf{a}\mathbf{b}} \mathbf{x} = \{ \mathbf{a}\mathbf{b}\mathbf{x} \} \\
K_{\mathbf{a}} \mathbf{x} =-\frac{1}{2} \{\mathbf{x}\mathbf{a}\mathbf{x}\} \nonumber
\end{eqnarray}
where
\begin{equation*}
  \{\mathbf{a}\mathbf{b}\mathbf{x}\}:= \mathbf{a}\circ (\mathbf{b}\circ \mathbf{x}) - \mathbf{b}\circ (\mathbf{a} \cdot \mathbf{x}) + (\mathbf{a}\circ \mathbf{b}) \circ \mathbf{x}
\end{equation*}
\begin{equation*}
  \mathbf{a},\mathbf{b},\mathbf{x} \in  J
\end{equation*}
and $\circ $ denotes the Jordan product. They satisfy the commutation relations
\begin{eqnarray}
[T_\mathbf{a}, K_\mathbf{b}]&= &M_{\mathbf{a}\mathbf{b}} \\ \nn
[M_{\mathbf{a}\mathbf{b}},T_\mathbf{c}]&=&T_{\{\mathbf{a}\mathbf{b}\mathbf{c}\}} \\ \nn
[M_{\mathbf{a}\mathbf{b}},K_\mathbf{c}] &=& K_{\{\mathbf{b}\mathbf{a}\mathbf{c}\}} \\ \nn
[M_{\mathbf{a}\mathbf{b}},M_{\mathbf{c}\mathbf{d}}]&=& M_{\{\mathbf{a}\mathbf{b}\mathbf{c}\}\mathbf{d}}- M_{\{\mathbf{b}\mathbf{a}\mathbf{d}\}\mathbf{c}} \nn
\end{eqnarray}
corresponding to the well-known Tits-Kantor-Koecher construction of Lie algebras from Jordan triple systems \cite{MR0321986,MR0146231,MR0228554}
We note that $M_{\mathbf{a}\mathbf{b}}$ can be written as
\begin{equation}
  M_{\mathbf{a}\mathbf{b}}= D_{\mathbf{a},\mathbf{b}} + L_{\mathbf{a}\cdot \mathbf{b}}
\end{equation}
where $D_{\mathbf{a},\mathbf{b}}$ generate the automorphism (rotation) group of $J$
\begin{equation*}
  D_{\mathbf{a},\mathbf{b}} \mathbf{x} = \mathbf{a}\circ
  (\mathbf{b}\circ \mathbf{x}) - \mathbf{b}\circ (\mathbf{a} \circ
  \mathbf{x})
\end{equation*}
and $L_{c}$ denotes multiplication by the element $c\in J$. The
generator $D$ is proportional to the multiplication operator by the
identity element of $J$.

Choosing a basis $e_I$ for the Jordan algebra such that an element
$\mathbf{x} \in J$ can be written as $\mathbf{x} = e_I q^I=\tilde{e}^I q_I$
one can write the generators of $\mathfrak{conf}(J)$ as differential
operators acting on the ``coordinates'' $q^I$. \footnote{ We should
note that there are , in general, two inequivalent actions of the
reduced structure group on the Jordan algebra and its conjugate. The
tilde refers to the conjugate basis such that $q_I p^I$ is invariant
under the action of the reduced structure group. For details on this
issue see \cite{Gunaydin:1992zh}.} These generators can be twisted
by a unitary character $\lambda$ and take the simple form
\begin{eqnarray}
  T_I & = &\frac{\partial}{\partial q^I}  \nonumber \\
  R^I_J& = &- \Lambda^{IK}_{JL} q^L \frac{\partial}{\partial q^K}  - \lambda  \delta^I_J \\
  K^I & =& \frac{1}{2} \Lambda^{IK}_{JL}  q^J q^L  \frac{\partial}{\partial q^K} + \lambda q^I \nonumber \\
\end{eqnarray}
where
\begin{equation*}
   \Lambda_{KL}^{IJ} := \delta_K^I \delta_L^J + \delta_L^I \delta^J_K - \frac{4}{3} C^{IJM} C_{KLM}
\end{equation*}
They satisfy the commutation relations
\begin{eqnarray}
& [ T_I, K^J ] =& - R^J_I \\
&[ R^J_I , T_K ] = &\Lambda_{IK}^{JL} T_L \\
&[R^J_I , K^K ] = &- \Lambda_{IK}^{JL} K^L
\end{eqnarray}
The generator of automorphism group are
\begin{equation}
A_{IJ} = R_I^J -R_J^I
\end{equation}
\setcounter{equation}0
\def\theequation{C.\arabic{equation}}

\section{ Quasiconformal realizations of Lie groups and Freudenthal triple systems }
\setcounter{equation}{0}
In this appendix we shall review the general theory of quasiconformal
realizations of noncompact groups over Freudenthal triple systems that
was given in \cite{Gunaydin:2000xr}.

Every simple Lie algebra $\mathfrak{g}$ of $G$ can be given a 5-graded
decomposition, determined by one of its generators $\Delta$, such that
grade $\pm 2$ subspaces are one dimensional:
\begin{eqnarray}
\mathfrak{g} = \mathfrak{g}^{-2} \oplus \mathfrak{g}^{-1} \oplus
       \mathfrak{g}^0 \oplus \mathfrak{g}^{+1}
     \oplus \mathfrak{g}^{+2} \,.
\end{eqnarray}
where
\begin{equation}
\mathfrak{g}^0 = \mathfrak{h} \oplus \Delta
\end{equation}
and
\begin{equation}
[\Delta, \mathfrak{t}]= m \mathfrak{t} \; \;\;\; \forall \mathfrak{t} \in \mathfrak{g}^m \;\;,\; m=0,\pm1,\pm2
\end{equation}
A simple Lie algebra with such a 5-graded decomposition can be
constructed over a Freudenthal triple system $\mathcal{F}$
\footnote{For $sl(2)$ the 5-grading degenerates into a 3-grading.}.
Freudenthal introduced these triple systems in his study of the
geometries associated with exceptional groups \cite{MR0063358}. A
Freudenthal triple system (FTS) is defined as a vector space
$\mathcal{F}$ equipped with a triple product $(X,Y,Z)$
\begin{equation}
(X,Y,Z) \in \mathcal{F} \;\;\;\;\;\;\;\; \forall \;\; X,Y,Z \in \mathcal{F}
\end{equation}
and a skew symmetric bilinear
form $\langle X,Y\rangle = -\langle Y,X\rangle$ such that the triple product satisfies the identities
\begin{eqnarray}
(X,Y,Z) &=&(Y,X,Z) +2\,\langle X,Y \rangle Z \,,\nonumber \\
(X,Y,Z) &=& (Z,Y,X) -2\,\langle X,Z \rangle Y \,,\nonumber \\
\langle (X,Y,Z), W \rangle  &=& \langle (X,W,Z),Y \rangle
                                -2\,\langle X, Z \rangle \langle Y ,W \rangle \,,\nonumber \\
(X,Y,(V,W,Z)) &=& (V,W,(X,Y,Z)
                             +((X,Y,V),W,Z) \nonumber \\
                          && {}+ (V,(Y,X,W),Z)  \,.\label{ftp56-rel}
\end{eqnarray}
The construction of the associated Lie algebra
$\mathfrak{g}(\mathcal{F})$ labels the generators belonging to
subspace $\mathfrak{g}^{+1}$ by the elements of $\mathcal{F}$
\begin{equation}
U_A \in \mathfrak{g}^{+1} \leftrightarrow \;\;\;\;\; A\in \mathcal{F}
\end{equation}
and by using the  involution , that reverses the grading, elements of
$\mathfrak{g}^{-1}$ can also be labeled by elements of $\mathcal{F}$
\begin{equation}
   \tilde{U}_A \in \mathfrak{g}^{-1} \leftrightarrow \;\;\;\;\; A\in \mathcal{F}
\end{equation}
Elements of $\mathfrak{g}^{\pm1}$ generate the Lie algebra
$\mathfrak{g}(\mathcal{F})$ by commutation, and hence the remaining
generators can be labelled by a pair of elements of $\mathcal{F}$
\begin{eqnarray}
 [U_A,\tilde{U}_B]         &\equiv & S_{AB} \;\; \in \mathfrak{g}^0            \\  \nn
 [U_A,U_B]            & \equiv & -K_{AB} \;\; \in \mathfrak{g}^2 \\  \nn
 [\tilde{U}_A,\tilde{U}_B] & \equiv & -\tilde{K}_{AB} \;\; \in \mathfrak{g}^{-2} \\ \nn
 \end{eqnarray}
Commutation relations are all determined by the Freudenthal triple
product $(A,B,C)$
\begin{eqnarray}
 [S_{AB},U_C]            &=& -U_{(A,B,C)}          \\ \nn
 [S_{AB},\tilde{U}_C]    &=& - \tilde{U}_{(B,A,C)} \\ \nn
 [K_{AB},\tilde{U}_C] &=& U_{(A,C,B)} - U_{(B,C,A)} \\ \nn
 [\tilde{K}_{AB},U_C] &=& \tilde{U}_{(B,C,A)} -\tilde{U}_{(A,C,B)} \\ \nn
 [S_{AB},S_{CD}]   &=& -S_{(A,B,C)D}
                            -S_{C (B,A,D)} \\ \nn
 [S_{AB},K_{CD}]         &=&  K_{A\FTP{C}{B}{D}}
                              - K_{A\FTP{D}{B}{C}} \\ \nn]
 [S_{AB},\tilde{K}_{CD}] &=&  \tilde{K}_{\FTP{D}{A}{C}B}
                              - \tilde{K}_{\FTP{C}{A}{D}B}\\  \nn
 [K_{AB},\tilde{K}_{CD}] &=&   S_{\FTP{B}{C}{A}D}
                                -S_{\FTP{A}{C}{B}D}
                                -S_{\FTP{B}{D}{A}C}
                                +S_{\FTP{A}{D}{B}C} \\ \nn
\end{eqnarray}
Since the grade $\pm 2$ subspaces are one dimensional one can write
\begin{equation}
  K_{AB} :=K_{\langle A , B\rangle}:=\langle A , B\rangle  K
\end{equation}
\begin{equation}
  \tilde{K}_{AB} :=\tilde{K}_{\langle A , B\rangle}:= \langle A, B\rangle  \tilde{K}
\end{equation}
Furthermore, the defining identities of the FTS imply that
\begin{equation}
S_{AB}-S_{BA} = -2 \langle A, B\rangle \Delta
\end{equation}
where $\Delta$ is the generator that determines the 5-grading
\begin{eqnarray}
  [\Delta, U_A]&=&U_A \\ \nn
  [\Delta, \tilde{U}_A ]&=& - \tilde{U}_A \\ \nn
  [\Delta , K]& = &2 K \\ \nn
  [\Delta, \tilde{K} ]& =& - 2 \tilde{K}
\end{eqnarray}
and generates the distinguished  $sl(2)$ subalgebra together with $K, \tilde{K}$
\begin{equation}
  [K, \tilde{K}] = -2 \Delta
\end{equation}
The 5-grading of $\mathfrak{g}$ is then given as
\begin{equation*}
   \mathfrak{g} = \tilde{K} \oplus \tilde{U}_A \oplus [S_{(AB)}+ \Delta ] \oplus U_A \oplus K
\end{equation*}
where
\begin{equation*}
  S_{(AB)} := \frac{1}{2} ( S_{AB} + S_{BA})
\end{equation*}
are the generators of $Aut(\mathcal{F})$ and commute with $\Delta$
\begin{equation}
  [\Delta, S_{(AB)} ]=0
\end{equation}
The remaining  non-zero commutators are
\begin{eqnarray}
  [ U_A, \tilde{U}_B ] &=& S_{(AB)} - \langle A, B \rangle \Delta \\ \nn
  [ K, \tilde{U}_A]&=& -2 \tilde{U}_A \\ \nn
  [ \tilde{K}, U_A]&=&  2 \tilde{U}_A \\ \nn
  [ S_{(AB)}, K]&=&0 \\ \nn
\end{eqnarray}
A quartic invariant $\mathcal{Q}_4$ can be defined over the FTS
$\mathcal{F}$ using the triple product and the bilinear form as
\begin{eqnarray}
\mathcal{Q}_4(X) := \frac{1}{48}\langle (X,X,X),X \rangle \label{e7-invariant}
\end{eqnarray}
which is invariant under the automorphism group $Aut(\mathcal{F})$ of
$\mathcal{F}$ generated by $S_{(AB)}$.

As was shown in \cite{Gunaydin:2000xr} one can realize the 5-graded
Lie algebra $\mathfrak{g}$ geometrically as a quasiconformal Lie
algebra over a vector space $\mathcal{T}$ coordinatized by the
elements $X$ of the FTS $\mathcal{F}$ and an extra single variable $x$
\cite{Gunaydin:2000xr,Gunaydin:2005zz}:
\begin{equation}
\begin{split}
  \begin{aligned}
      K\left(X\right) &= 0 \\
      K\left(x\right) &= 2\,
  \end{aligned}
  & \quad
  \begin{aligned}
     U_A \left(X\right) &= A \\
     U_A\left(x\right) &= \left< A, X\right>
  \end{aligned}
   \quad
   \begin{aligned}
      S_{AB}\left(X\right) &= \left( A, B, X\right) \\
      S_{AB}\left(x\right) &= 2 \left< A, B\right> x
   \end{aligned}
 \\
 &\begin{aligned}
    \Tilde{U}_A\left(X\right) &= \frac{1}{2} \left(X, A, X\right) - A x \\
    \Tilde{U}_A\left(x\right) &= -\frac{1}{6} \left< \left(X, X, X\right), A \right> + \left< X, A\right> x
 \end{aligned}
 \\
 &\begin{aligned}
    \Tilde{K}\left(X\right) &= -\frac{1}{6} \,  \left(X,X,X\right) +  X x \\
    \Tilde{K}\left(x\right) &= \frac{1}{6} \,  \left< \left(X, X, X\right), X \right> + 2\,  \, x^2
 \end{aligned}
\end{split}
\end{equation}
The geometric nature of quasiconformal actions is made manifest by
first defining a quartic norm over the space $\mathcal{T}$ as
\begin{equation}
\cN_4(\cX) := \mathcal{Q}_4(X) - x^2
\end{equation}
where $\mathcal{Q}_4(X)$ is the quartic invariant of $\mathcal{F}$ and
then a ``distance'' function between any two points $\cX=(X,x)$ and
$\cY=(Y,y) $ in $\mathcal{T}$ as
\begin{equation}
d(\cX,\cY):= \cN_4(\gd(\cX,\cY)
\end{equation}
where  $\gd(\cX,\cY)$ is the ``symplectic'' difference of two  vectors $\cX $ and $\cY$ :
\begin{equation*}
  \gd(\cX,\cY) := (X-Y,x-y+\langle X, Y \rangle )
\end{equation*}
One can then show that the light-like separations with respect to this
quartic distance function
\begin{equation*}
  d(\cX,\cY)=0
\end{equation*}
is left invariant under the action of quasiconformal group
\cite{Gunaydin:2000xr}.
\bibliography{QCGTWISTOR_HEPTH2}
\bibliographystyle{utphys}
\end{document}